\documentstyle[preprint,prc,aps,tighten,floats,epsf]{revtex}

\def\omegaV{V}
\def\omegav{{\rm v}}
\def\gomega{g_\omegav}
\def\momega{m_\omegav}
\def\fomega{f_\omegav}
\def\fpi{f_\pi}
\def\rhoB{\rho_{\scriptscriptstyle\rm B}}
\def\Mstar{M^*}

\begin{document}

%

\preprint{\vbox{\hfill IU/NTC\ \ 96--09\\
                \null\hfill OSU--96--612}}

\title{A Chiral Effective Lagrangian For Nuclei}

\author{R. J. Furnstahl}
\address{Department of Physics\\
         The Ohio State University,\ \ Columbus,~Ohio\ \ 43210}
\author{Brian D. Serot}
\address{Physics Department and Nuclear Theory Center\\
         Indiana University,\ \ Bloomington,~Indiana\ \ 47405}
\author{Hua-Bin Tang%
  \footnote{Present address: School of Physics and Astronomy,
     University of Minnesota, Minneapolis, MN\ \ 55455.}}
\address{Department of Physics\\
         The Ohio State University,\ \ Columbus,~Ohio\ \ 43210}
\date{August, 1996}
\maketitle

\begin{abstract}
An effective hadronic lagrangian consistent with the symmetries of 
quantum chromodynamics and intended for applications to finite-density 
systems is constructed.
The degrees of freedom are (valence) nucleons, pions, and the low-lying
non-Goldstone bosons, which account for the intermediate-range
nucleon--nucleon interactions and conveniently describe the nonvanishing
expectation values of nucleon bilinears.
Chiral symmetry is realized nonlinearly, with a light scalar meson
included as a chiral singlet to describe the mid-range
nucleon--nucleon attraction.
The low-energy electromagnetic structure of the nucleon is
described within the theory using vector-meson dominance, so
that external form factors are not needed.
The effective lagrangian is expanded in powers of the fields and
their derivatives, with the terms organized using Georgi's
``naive dimensional analysis''.
Results are presented for finite nuclei and nuclear matter 
at one-baryon-loop order, using the single-nucleon structure
determined within the model.
Parameters obtained from fits to nuclear properties show that naive 
dimensional analysis is a useful principle and that a truncation
of the effective lagrangian at the first few powers of the fields
and their derivatives is justified.
\end{abstract}

\vspace{20pt}
%


\section{Introduction}

Quantum chromodynamics (QCD) 
is generally accepted as the underlying theory of the
strong interaction. 
At low energies relevant to nuclear physics, however,
nucleons and mesons are convenient and efficient degrees of freedom.
In particular, 
relativistic field theories of hadrons, called quantum
hadrodynamics (QHD), have been quite successful in describing the bulk and 
single-particle properties of nuclei and nuclear matter in the mean-field
and Dirac--Brueckner--Hartree--Fock 
approximations\cite{SEROT86,SEROT92}.

QHD studies based on renormalizable models, however, have encountered 
difficulties due to large effects from loop integrals that incorporate
the dynamics of the quantum vacuum\cite{PERRY87,FURNSTAHL88,FURNSTAHL89}.
On the other hand, 
the ``modern'' approach to renormalization
\cite{LEPAGE89,POLCHINSKI92,GEORGI94,BALL94,WEINBERG95},
which makes sense of
effective, ``cutoff'' theories\footnote{In this work, ``cutoff'' means a 
regulator that maintains the appropriate symmetries.}
with  low-energy, composite degrees
of freedom, provides an alternative. 
When composite degrees of  freedom are used, the structure of
the particles is described with increasing detail by
including more and more {\it nonrenormalizable\/} interactions
in a derivative expansion\cite{LEPAGE89}.
We seek to merge the successful phenomenology of the QHD approach
to nuclei with  
the modern viewpoint of
effective field theories.

The present work is an extension of Refs.~\cite{FS,vacuum,bodmer},
which explored some aspects of incorporating QCD symmetries into effective
hadronic models.
Here we build a more complete effective lagrangian and
apply it to nuclei at the one-baryon-loop level.
Our two main goals are to include effects of compositeness, such as
single-nucleon electromagnetic structure, in a unified, self-consistent
framework and to test whether ``naive dimensional analysis'' (NDA)
and naturalness (see below) are useful concepts for effective
field theories of nuclei. 
Eventually, we will need to demonstrate a systematic expansion scheme
that includes corrections beyond one-loop order.
Nevertheless, Ref.~\cite{bodmer} describes how a general energy 
functional with classical meson fields can implicitly incorporate 
higher-order contributions.
We therefore expect that higher-order many-body effects can be absorbed
into the coefficients of our one-loop energy functional when we determine
parameters by fitting to nuclear observables.
We return to this point later in this section.

Important  physical constraints on the effective lagrangian
are established by maintaining the symmetries of QCD. 
These include Lorentz invariance, parity invariance, 
electromagnetic gauge invariance, 
isospin, and chiral symmetry.
These symmetries constrain the lagrangian most directly by restricting
the form of possible interaction terms.
In some cases, the magnitude of (or relations between) parameters may
also be restricted.

Chiral symmetry has been known to play an important role 
in hadronic physics at low energies since the early days of current 
algebra \cite{ADLER68,ALFARO73}; it is exploited 
systematically in terms of effective fields in the framework known 
as chiral perturbation theory (ChPT)\cite{WEINBERG79,DASHEN69,GASSER84}. 
There is also a long history of attempts to unite  relativistic 
mean-field phenomenology based on hadrons with manifest chiral symmetry.
In particular, it has been tempting to build upon the linear 
sigma model \cite{GELLMANN60,LEE72,LEE74}, 
with a light scalar meson playing a dual role as the chiral 
partner of the pion {\it and\/} the mediator of the intermediate-range
nucleon--nucleon (NN) attraction. 
The generic failure of this type of model has been shown
in Refs.~\cite{FS,bodmer}.%
\footnote{The failure of models based on the linear sigma model
lies more with the assumed ``Mexican-hat'' potential than with the
linear realization of chiral symmetry.  Successful mean-field models
with a linear realization and a light scalar meson have been constructed
using a logarithmic potential (see Refs.~\protect\cite{HEIDE94,CARTER96}).}
However, a light scalar is completely consistent with a {\it nonlinear\/}
realization of chiral symmetry \cite{vacuum,bodmer}.
Moreover, in a nonlinear realization, one need not assume that the
hadrons are grouped into particular chiral multiplets \cite{WEINBERG68}.

The phenomenological success of  ChPT establishes the dominant
contributions of the vector resonances to the low-energy constants in
the pion sector\cite{ECKER89a,ECKER89b,DONOGHUE89}. 
In particular, a tree-level effective lagrangian with vector mesons is 
found to be essentially equivalent to ChPT at one-meson-loop order. 
In the  baryon sector,  
a satisfactory description of the nucleon form
factors from vector-meson dominance (VMD) \cite{VMDa,VMDb,VMDc} 
alone has not been found\cite{BROWN86}.
However, we {\it can\/} describe the long-range
parts of the single-nucleon form factors (that is, magnetic moments and 
mean-square radii) 
using an effective lagrangian that includes vector mesons
and direct couplings of photons to nucleons.
Our approach is analogous to that implemented long ago in 
Ref.~\cite{IACHELLO73}.

In virtually all previous QHD calculations of finite nuclei, the
single-nucleon electromagnetic structure was incorporated ``by hand.''
After the nuclear structure calculation was completed and ``point-nucleon''
densities were obtained, they were convoluted with empirical 
single-nucleon form factors to compute the charge density.
The results were
typically insensitive to the nature of the form factor, as long as the 
nucleon charge radius was accurate.
In the present work, we do not need external form factors.
The large oscillations in the point-nucleon charge density of a nucleus
are significantly reduced by virtual mesons inherent in the model, and 
the low-momentum parts of the charge form factors of various nuclei are 
well reproduced.
Thus we have a {\it unified\/} framework.

To achieve a systematic framework for calculating nuclei, however,
we must develop an expansion scheme.
ChPT is an expansion in powers of  small external momenta and the
pion mass. 
Weinberg\cite{WEINBERG90} applies this small-momentum
expansion further to a calculation of the $N$-nucleon 
``effective potential'', defined as the sum of all connected diagrams 
for the $S$ matrix (in time-ordered perturbation theory)
without $N$-nucleon intermediate states. 
On the other hand, an expansion of the NN
interaction kernel 
in terms of intermediate states of successively shorter range
provides a successful and economical way to accurately
describe NN scattering data \cite{MACHLEIDT}.
Here we  combine the ideas of the small-momentum expansion and
the intermediate-state expansion by including the low-lying 
{\it non-Goldstone bosons\/} into the low-energy effective lagrangian.
These bosons account for the intermediate-range interactions
and conveniently describe the nonvanishing expectation values
of nuclear bilinears such as $\overline{N}N$ and 
$\overline{N}\gamma_\mu N$.

There are well-defined resonances in the vector channels, and thus it is 
natural to identify them with fields in the effective lagrangian; in contrast,
the dynamics in the scalar channel is more complicated. 
Numerous calculations\cite{JACKSON75,DURSO77,DURSO80,LIN89,LIN90} show that
correlated two-pion exchange leads to a distribution of strength
in the isoscalar-scalar channel at center-of-mass pion energies of roughly 
500 to 600~MeV, even in the absence of a low-energy resonance.
We therefore introduce an isoscalar-scalar meson as an {\em effective\/} 
degree of freedom to simulate this strong interaction, just like the 
non-Goldstone bosons in the vector channels.
Such a scalar, with a mass of 500 to 600~MeV, is essential in
one-boson-exchange models of the NN potential
\cite{MACHLEIDT} and in relativistic mean-field models of 
nuclei\cite{SEROT86}.
We emphasize that at the level of approximation used in this paper, 
the scalar field can be introduced without loss of generality.
(For example, it can be viewed as a way to remove terms with
more than two nucleon fields from a general pion-nucleon lagrangian.)
A deeper understanding of the NN interaction in the scalar channel
is an important pursuit, but is not needed for our purposes here. 


We shall expand our effective hadronic lagrangian in powers 
of the fields and their derivatives, and organize the terms using 
``naive dimensional analysis'' (NDA)\cite{GEORGI93}. 
The NDA framework is used to identify  the dimensional factors of any term. 
After these dimensional factors and some appropriate
counting factors are extracted, the remaining {\em dimensionless\/}
coefficients are all assumed to be of order unity.
This is the so-called naturalness assumption.
If naturalness is valid, the effective lagrangian can be truncated
at a given order with a reasonable bound on the truncation error for
physical observables.
Thus we have a controlled expansion, at least at the tree level.

Evidently, for an expansion in powers of the fields and their derivatives 
to be useful, the naturalness assumption must hold.
Implicit in this assumption is that all 
the short-distance physics is
incorporated into the coefficients of the effective lagrangian; otherwise,
the truncation error will not be reliably given by NDA power counting
when loops are included.
Thus, just as in ChPT or in heavy-baryon theories, the nucleon vacuum loops 
and non-Goldstone-boson loops will be ``integrated out'', with their 
effects buried implicitly in the coefficients.%
\footnote{We note that the one-nucleon-loop vacuum
contributions to the coefficients in renormalizable mean-field models 
(included in the so-called relativistic Hartree approximation,
or RHA \cite{SEROT86}) are {\em not\/} natural\cite{FSTprepare}.
In light of our current results in favor of natural coefficients, we view
the unnatural RHA coefficients as evidence that the vacuum is not treated
adequately.}
Nevertheless, the nucleon and non-Goldstone-boson fields are still needed 
in the lagrangian to account for the valence nucleons, to treat the mean 
fields conveniently, and to apply the usual many-body and field-theoretic
techniques. 
Any contributions from vacuum nucleon  or non-Goldstone-boson loops 
that arise in many-body calculations can formally be cancelled by 
the appropriate counterterms.
These cancellations can always be achieved, since 
{\it all possible interaction terms consistent 
with the symmetries are already included in the effective lagrangian.}
On the other hand, long-distance, finite-density effects
should be calculated explicitly in a systematic application 
of the effective lagrangian.

In principle, the naturalness assumption can be verified by  deriving the
effective lagrangian from QCD. 
Since this is not yet feasible, one must fit the unknown parameters 
to experimental data. 
At finite density, however, the Fermi momentum enters as a new scale. 
This complicates the situation, because a treatment of the effective 
lagrangian at the level of classical meson fields and valence nucleons
(the Dirac--Hartree approximation) implies that several different physical
effects are implicitly contained in the parameters.
Not only are short-distance contributions from meson and nucleon vacuum
loops absorbed, but long-distance, many-body effects from well-known 
ladder and ring diagrams are included approximately as well \cite{bodmer}.
Thus the usefulness of NDA, the naturalness assumption,  and the
expansion in powers of the fields and their derivatives requires 
further examination when these ideas are applied to problems 
at finite density.
It is not obvious that an effective lagrangian with natural coefficients
that include only short-distance physics will remain natural
in a one-loop calculation that accounts for higher-order many-body effects 
by adjustments of the coefficients.
Here we rely on fits to the properties of finite nuclei to determine whether 
the coefficients are natural and our expansion of the lagrangian is useful.
(Friar and collaborators have recently investigated naturalness as 
exhibited by a mean-field point-coupling model\cite{FRIAR95}.)
An important topic for future study is to include the many-body dynamics
explicitly, so as to disentangle it from the short-range and vacuum
dynamics already implicit in the parameters of the low-energy effective
lagrangian.


The interaction terms of the nucleons and the non-Goldstone bosons
can be quite complicated, and different forms have been used in various 
models in the literature \cite{ZIMANYI90,DELFINO94,LENSKE95}. 
Some of these models are actually equivalent if one uses the
freedom of field redefinitions\cite{GEORGI91}, which leave the on-shell
$S$ matrix invariant\cite{COLEMAN69,CALLAN69,BANDO88}.%
\footnote{We {\em assume\/} that this is also sufficient for finite-density
observables, but we know of no proof.}
We discuss how to take advantage of this freedom to simplify the interaction 
terms between the nucleons and the non-Goldstone bosons. 
As a result,  the lagrangian can be conveniently
written in a ``standard form'', in which one mainly has
the traditional Yukawa interactions.

The rest of the paper is organized as follows.
In Sec.~II, we discuss the construction of an effective hadronic lagrangian 
consistent with chiral symmetry, gauge invariance, and Lorentz and parity
invariance. 
Chiral symmetry is realized in the standard nonlinear form of Callan, 
Coleman, Wess, and Zumino (CCWZ)\cite{CALLAN69}.
Neutral scalar and vector mesons are introduced to efficiently describe
the mid- and short-range parts of the NN interaction, and the role of
broken scale invariance in QCD is considered.
Georgi's naive dimensional analysis and field redefinitions are also
discussed, with the
new feature of off-shell non-Goldstone bosons. 
In Sec.~III, we show how the electromagnetic structure of the pion
and the nucleon can be described in the low-energy regime.
Applications to nuclear matter and finite nuclei at the Dirac--Hartree level
are presented in Sec.~IV, and the results are shown in Sec.~V. 
A discussion and summary is given in Sec.~VI.

\section{Constructing the Lagrangian}

We begin this section by showing how we implement nonlinear chiral symmetry
and electromagnetic gauge invariance using nucleons, Goldstone pions, and
non-Goldstone isovector mesons.
Next, we discuss the introduction of neutral vector and scalar mesons that
are used to provide an efficient description of the short- and mid-range
nucleon--nucleon interaction.
To use the resulting lagrangian, which in principle has an infinite number
of terms, we must have some systematic truncation procedure, and we develop
this using the ideas of naive dimensional analysis and naturalness, together 
with some observations on the sizes of meson mean fields in nuclei.

\subsection{Chiral Symmetry}

We illustrate a nonlinear realization of chiral symmetry\cite{CALLAN69}
using a system that consists of rho mesons, pions, and nucleons only. 
We construct a general effective lagrangian that respects chiral 
symmetry, Lorentz invariance, and parity conservation. 
The role of vector-meson dominance (VMD) is also discussed.
We restrict ourselves to the chiral limit with massless pions 
and postpone the generalization
to a finite pion mass for future study.

The Goldstone pion fields $\pi^a(x)$,
with $a=1,\, 2,\, {\rm and}\, 3\,$,
form an isovector, which can be considered as the phase of 
a chiral rotation of the identity matrix in isospin space:
\begin{equation}
 \xi {\bf 1} \xi \equiv U(x) =
     \exp (i\pi(x)/ f_{\pi})
        {\bf 1}    \exp (i \pi(x)/ f_{\pi})
               \ . \label{eq:U-def}
\end{equation}
  Here $f_{\pi} \approx 93\,\mbox{MeV}$ 
  is the pion-decay constant and the pion field is compactly
written as $\pi(x) \equiv \bbox{\pi}(x)\, \bbox{\cdot}\,{1\over 2}\bbox{\tau}$,
with $\tau^a$ being Pauli matrices.
The matrix $U$ is the standard exponential representation and
the ``square root'' representation in terms of $\xi$  is
particularly convenient for including heavy fields in the chiral
lagrangian.
The isospinor nucleon field is represented by a column matrix
\begin{equation}
N(x)=\left (\,
     \begin{array}{c} 
         p(x) \\ 
         n(x)
      \end{array} \right )\ ,
\end{equation}
where  $p(x)$ and $n(x)$ are the proton and neutron fields
respectively. 
The rho fields $\rho^a_\mu(x)$ also form an isovector;  we use the 
notation $\rho_\mu(x) \equiv \bbox{\rho}_\mu (x) 
\,\bbox{\cdot}\,{1\over 2}\bbox{\tau}$.
(See Refs.~\cite{ECKER89b,BIJNENS95,BORASOY95} for discussions of
the formal equivalence between the vector and the tensor
 formulations for spin-1 fields.)

   Following CCWZ\cite{CALLAN69}, we define a  nonlinear realization of 
the chiral group $SU(2)_{\rm L}\otimes SU(2)_{\rm R}$ such that,
for arbitrary global matrices $L \in SU(2)_{\rm L}$ and $R \in SU(2)_{\rm R}$,
we have the mapping
\begin{equation}
L\otimes R:\ \ \ (\xi, \rho_\mu, N)\longrightarrow 
        (\xi', \rho'_\mu, N')   
          \ ,     \label{eq:nonlr}
\end{equation}
where 
\begin{eqnarray}
\xi'(x) &=& L \xi(x) h^{\dagger}(x) = h(x) \xi(x) R^{\dagger}
               \ , \label{eq:Xitrans} \\[4pt]
\rho'_\mu(x) &=& h(x) \rho_\mu (x) h^{\dagger}(x)
               \ , \label{eq:Rhotrans} \\[4pt]
 N'(x) &=& h(x)N(x)  \ .       \label{eq:Ntrans}
\end{eqnarray}
As usual, the matrix $U$ transforms as $U'(x) = LU(x)R^{\dagger}$.
The second equality in Eq.~(\ref{eq:Xitrans}) defines
$h(x)$ implicitly as a function of $L$, $R$, and the local pion fields:
$h(x)=h(\bbox{\pi}(x),L,R)$.%
\footnote{We can express $h$ in terms of $L$, $R$, and $U$ as
$h(x) = \sqrt{{U'}^{\dagger}(x)}L\sqrt{\vphantom{{U'}^{\dagger}}U(x)} =
\sqrt{RU^{\dagger}(x)L^{\dagger}}\,
L\,\sqrt{\vphantom{U^{\dagger}}U(x)}$.  Given
the decomposition \cite{COLEMAN69}
$L=\exp(i\bbox{\alpha\cdot\tau})\exp(i\bbox{\beta\cdot\tau})$,
$R=\exp(-i\bbox{\alpha\cdot\tau})\exp(i\bbox{\beta\cdot\tau})$,
and  $h=\exp(i\bbox{\gamma\cdot\tau})$,
with $\bbox{\alpha}$, $\bbox{\beta}$, and $\bbox{\gamma}$ real,
the infinitesimal expansion of $h$ is  
$\bbox{\gamma} = \bbox{\beta} - (\bbox{\alpha}\times\bbox{\pi})/2\fpi 
  + O(\bbox{\alpha}^2, \bbox{\beta}^2, \bbox{\pi}^2)$ \cite{BERNARD95}.}
Under parity, which is defined to be 
\begin{equation}
{\cal P}:\ \ \ L \rightarrow R \ , \ \ R \rightarrow L \ ,
\label{eq:parity}
\end{equation}
the pion field transforms as a pseudoscalar: 
$\xi(t,{\bf x}) \rightarrow \xi^\dagger(t,{\bf -x})$.
The existence of the parity automorphism (\ref{eq:parity}) allows the
simple decomposition  $U = \xi\xi$ in Eq.~(\ref{eq:U-def}) 
\cite{CALLAN69}.
It follows from Eq.~(\ref{eq:Xitrans}) that 
$h(x)$ is invariant under parity, that is, $h(x)\in SU(2)_{\rm V}$, with
$ SU(2)_{\rm V}$ the unbroken vector  subgroup
of $SU(2)_{\rm L} \otimes SU(2)_{\rm R}$.
Eqs.~(\ref{eq:Rhotrans}) and (\ref{eq:Ntrans}) 
ensure that the rho and nucleon transform linearly under
$SU(2)_{\rm V}$ in accordance with their isospins. 
Note that the matrix $h(x)$  becomes global only for 
 $SU(2)_{\rm V}$ transformations,
in which case $h=L=R$.

 
It is straightforward to show that Eqs.~(\ref{eq:nonlr})
to (\ref{eq:Ntrans}) indeed yield a realization
of the chiral group \cite{COLEMAN69}. We emphasize that there is no local
gauge freedom involved, since there are no arbitrary
local gauge functions in the transformations.
The nonlinear realization
implies that chiral rotations involve the absorption
and emission of pions, as expected from a spontaneously broken
symmetry\cite{WEINBERG68,GEORGI84}.
For example, while vector transformations mix the proton with the
neutron, axial transformations mix the proton with states of nucleons
plus zero-momentum pions. 

To complete the building blocks of the chiral effective lagrangian,
we define an axial vector field
$a_{\mu}(x)$ and a polar vector field $v_{\mu}(x)$ by
\begin{eqnarray}
a_{\mu}     & \equiv & -{i \over 2}(\xi^{\dagger} \partial_{\mu} \xi -
    \xi \partial_{\mu}\xi^{\dagger} )
          = a_{\mu}^\dagger \ ,\label{eq:adef}\\[4pt]
v_{\mu}     & \equiv & 
           -{i \over 2}(\xi^{\dagger} \partial_{\mu} \xi +
    \xi \partial_{\mu}\xi^{\dagger} )
         = v_{\mu} ^\dagger
            \ , \label{eq:vdef}
\end{eqnarray}
both of which contain one derivative. 
Under a chiral transformation, we find
\begin{eqnarray}
v_\mu   &\rightarrow  &   h v_\mu h^{\dagger}
                   -ih\partial_\mu h^{\dagger}
               \ , \\[3pt]
a_\mu   &\rightarrow &    h a_\mu h^{\dagger} \ .
\end{eqnarray}
Note that to leading order in derivatives,
\begin{equation}
 a_{\mu}={1\over \fpi}\partial_{\mu}\pi+ \cdots \ ,\ \ \
 v_{\mu}=-{i\over 2\fpi^2}[\pi,\partial_{\mu}\pi] + \cdots \ .
   \label{eq:leading}
\end{equation}
To maintain chiral invariance, instead of using an ordinary
derivative $\partial^\mu$ acting on a
rho or nucleon field, we should use a covariant derivative $D^\mu$.
The chirally covariant derivative of the rho field is
\begin{equation}
D_\mu\rho_\nu = \partial_\mu\rho_\nu + i [v_\mu , \rho_\nu] \ .
  \label{eq:rhocovar}
\end{equation}
We also define the covariant tensors 
\begin{eqnarray}
v_{\mu\nu} &=& \partial_{\mu} v_{\nu} -\partial_{\nu}
      v_{\mu} + i [v_{\mu}, v_{\nu}] 
       = - i [a_\mu, a_\nu]   \ , \\[4pt]
\rho_{\mu\nu}&=& D_\mu\rho_\nu - D_\nu\rho_\mu + i g [\rho_\mu,\rho_\nu] \ .
                         \label{eq:rhomunu}
\end{eqnarray}
The term proportional to $[\rho_\mu,\rho_\nu]$
on the right-hand side of Eq.~(\ref{eq:rhomunu}) vanishes
at the mean-field level, so the coupling $g$ will be left unspecified.

To write a general effective lagrangian, we need an organizational scheme
for the interaction terms. We organize the lagrangian in increasing powers
of the fields and their derivatives. While we do not consistently
include loops in this investigation, we assign
to each interaction term a size of  the order
of $k^{\nu}$, with $k$ a generic small momentum, and
\begin{equation}
\nu = d+{n\over 2} + b  \ ,   \label{eq:order}
\end{equation}
where $d$  is the number  of derivatives,
$n$ the number of nucleon fields, and $b$
the number of  non-Goldstone boson fields in the interaction term.
The first two terms in Eq.~(\ref{eq:order}) are suggested by 
Weinberg's work \cite{WEINBERG90}.
The last term is a generalization that arises because 
a non-Goldstone boson couples to {\it two} nucleon fields.
Equation~(\ref{eq:order}) is also consistent with finite-density applications 
when the density is not too much higher than the nuclear-matter equilibrium
density, as we will see in Sec.~II.D.

Taking into account
chiral symmetry, Lorentz invariance, and parity conservation, we
may write the lagrangian through quartic order ($\nu \leq 4$) as
\begin{eqnarray}
{\cal L}_1(x) &=&
         \overline N \Big [i\gamma^{\mu} (\partial_{\mu}
               + i v_\mu +i g_\rho \rho_\mu)
             +g_{\rm \scriptscriptstyle A}
             \gamma^{\mu}\gamma_5a_{\mu}
              -M \Big ]N \nonumber \\
         & & -{f_\rho g_\rho\over 4M}\overline N
             \rho_{\mu\nu}    \sigma^{\mu\nu} N
           - {\kappa_\pi\over M} \overline{N} 
              v_{\mu\nu}\sigma^{\mu\nu}  N
                           \nonumber \\
         & &  
            -{1\over 2}{\rm tr}(\rho_{\mu\nu}\rho^{\mu\nu})
               + m_\rho^2 {\rm tr}(\rho_\mu \rho^\mu) 
            -g_{\rho \pi\pi}{2\fpi^2 \over m_\rho^2}
              {\rm tr}\,(\rho_{\mu\nu}v^{\mu\nu})
                  \nonumber  \\
      & &
             + \fpi^2{\rm tr}(a_\mu a^\mu) 
             + {4\beta_\pi\over M} \overline{N} N 
               {\rm tr}\, (a_\mu a^\mu)
             + l_1[{\rm tr}(a_\mu a^\mu)]^2
             + l_2 {\rm tr}(v_{\mu\nu} v^{\mu\nu})
                  \ , \label{eq:Lag1}
\end{eqnarray}
where $\sigma_{\mu\nu}=i[\gamma_{\mu},\gamma_{\nu}]/2$,
$g_{\rm \scriptscriptstyle A}\approx 1.26$ is the axial coupling
constant, $M$ is the nucleon mass, $g_\rho$ is the $\rho N N$ coupling,
$f_\rho$ is the so-called tensor coupling,
$g_{\rho \pi\pi}$ is the $\rho \pi\pi$ coupling,
and $\beta_\pi$, $\kappa_\pi$, $l_1$, and $l_2$ are couplings for 
higher-order $\pi N$ and $\pi\pi$ interactions.
The pion couplings can be determined from $\pi N$ and $\pi\pi$
scattering\cite{GASSER84}. We have omitted terms 
with higher powers  of the
rho fields, which have small expectation values in finite nuclei, 
and four- or more-nucleon ``contact'' terms, which can be represented
by  appropriate powers of other meson fields, as discussed shortly.
A third-order ($\nu=3$) term of the form 
$\overline{N} N{\rm tr}(\rho_\mu \rho^\mu)$ will be included by coupling 
a scalar to two rho fields in Sec.~II.E.
The antisymmetry of rho meson field tensor $\rho_{\mu\nu}$ 
guarantees three independent degrees of freedom, because 
there is no momentum
conjugate to the time component $\rho_0$, and $\rho_0$ 
satisfies an equation of constraint.

 
Chiral symmetry permits isoscalar-scalar and isoscalar-vector terms 
like $ (\overline{N} N)^2$ and $ (\overline{N}\gamma_\mu N)^2$.
However, analyses of NN scattering with one-boson 
exchange potentials
\cite{MACHLEIDT} shows that the low-lying
mesons are sufficient for describing low-energy many-nucleon systems and
that they dominate the NN interaction.
One may then expect that contact interactions of 
four (or more) fermions can be removed in favor of couplings to massive
non-Goldstone bosons.
The following example shows that this should be a good approximation.

    Consider an additional isovector-vector contact term 
$ [\overline{N} (\bbox{\tau}/2)\gamma_\mu N]^2$. The contribution of
this term can be absorbed
completely into a renormalization of $g_\rho$
as follows.  If we had constructed ${\cal L}$ by
retaining several isovector-vector non-Goldstone bosons, we could 
eliminate them using their field equations to obtain
\begin{equation}
 -{1\over 2}\Big(
      {{{g'_\rho}}^2  \over m_\rho^2} +
        {g_1^2 \over m_1^2} + \cdots \Big)
        [\overline{N}(\bbox{\tau}/2)\gamma_\mu N]^2
  - {1\over 2}\Big(
      {{g'_\rho}^2  \over m_\rho^4} +
        {g_1^2 \over m_1^4} + \cdots \Big)
        \{\partial_\mu[\overline{N}(\bbox{\tau}/2)\gamma_\nu N]\}^2
        + \cdots
                  \ , \label{eq:rhoexp}
\end{equation}
where $g'_\rho$ is the $\rho NN$ coupling when the heavier mesons are
retained, with couplings to nucleons given by $g_1$, \dots\ and
masses given by $m_1$, \dots\ .
Evidently, the effects of the heavier particles can be absorbed into
a redefinition of the $\rho N N$ coupling:
\begin{equation}
 g_\rho^2 = {g'_\rho}^2 + {m_\rho^2 \over m_1^2}\, g_1^2 + \cdots \ .
\end{equation}
If we approximate the second term in Eq.~(\ref{eq:rhoexp}) as
\begin{equation}
   \Big( -{1\over 2}\, {g_\rho^2 \over m_\rho^4} \Big)
   \{\partial_\mu[\overline{N}(\bbox{\tau}/2)\gamma_\nu N]\}^2 \ ,
\end{equation}
the leading error is given by the following term:
\begin{equation}
  {g_1^2\over g_\rho^2}\, {m_\rho^2(m_1^2-m_\rho^2) \over m_1^4}
   \Big ( -{1\over 2}\, {g_\rho^2 \over m_\rho^4} \Big)
   \{\partial_\mu[\overline{N}(\bbox{\tau}/2)\gamma_\nu N]\}^2 \ .
\end{equation}
For a typical resonance in the isovector-vector channel
\cite{PDT}, we have $m_1=1450\,\mbox{MeV}$ and 
$m_\rho^2(m_1^2-m_\rho^2)/m_1^4 \approx 1/5$. 
If $g_1^2 /g_\rho^2$ is roughly $1/4$, the error is about
$5\%$ of the already small second term. 
One should be able to test the sensitivity to these 
errors by allowing the non-Goldstone boson masses to vary in the fit
to nuclear properties.

If we were to assume universal VMD in Eq.~(\ref{eq:Lag1}), then 
\begin{equation}
g_{\rho \pi\pi}=g_{\rho} \ , \qquad\qquad \kappa_\pi = {f_\rho\over 4} \ ,
\end{equation}
and $g=g_\rho$ in Eq.~(\ref{eq:rhomunu}).
If we further assume
the Kawarabayashi--Suzuki--Riazuddin--Fayyazuddin (KSRF)
\cite{KSRFa,KSRFb} relation
\begin{eqnarray}
g_{\rho \pi\pi} = {m_{\rho}^2 \over 2 g_\rho f_{\pi}^2} \ ,
\end{eqnarray}
then we may redefine the rho field to be $\rho_\mu + v_\mu/g_\rho$ and the
covariant field tensor to be $\rho_{\mu\nu} + v_{\mu\nu}/g_\rho$,
and recover the lagrangian of Weinberg\cite{WEINBERG68} or
Bando {\em et~al}.\cite{BANDO85}.

\subsection{Gauge Invariance and the Inclusion of the $\omega$ Meson}

    In finite nuclei, electromagnetism also plays an important role.
We may include electromagnetic gauge invariance in our formalism in a
straightforward manner by noticing how the pion, rho, and
nucleon fields transform under the local $U(1)_{Q}$ symmetry.
Here the electric charge $Q$,   the third
component of the isospin $T_3$, and the hypercharge $Y$ are related by
$Q=T_3+Y/2$.


Under  $U(1)_{Q}$,  the electromagnetic field $A_{\mu}$ transforms
in the familiar way
\begin{equation}
A_{\mu} \rightarrow A_{\mu}- {1\over e}\,\partial_{\mu}\alpha(x)
               \ .
\end{equation}
The pion, rho, and nucleon fields transform under
the local $U(1)_{Q}$ rotation just as in
Eqs.~(\ref{eq:Xitrans}) to (\ref{eq:Ntrans}), with both $L$ and $R$ set
equal to 
\begin{equation}
 h(x) = \exp \Big[ i\alpha(x)\Big({\tau_3+Y\over 2}\Big)\Big] \ ,
\end{equation}
where $Y=0$ for the pion and the rho, and $Y=1$
for the nucleon. 
Indeed, the resulting transformations for these
hadrons are just as expected from their electric charges. 
That is,
\begin{eqnarray}
  (\pi^0,\ \pi^\pm) &\rightarrow &(\pi^0,\ 
       e^{\pm i\alpha}\pi^\pm)
         \ ,         \\[5pt]
(\rho_\mu^0,\ \rho_\mu^\pm) &\rightarrow &
      (\rho_\mu^0,\   e^{\pm i\alpha}\rho_\mu^\pm )
         \ ,        \\[5pt]
 ( n,\ p)       &\rightarrow & (n,\
       e^{+i\alpha}p )  \  .
\end{eqnarray}

Electromagnetism breaks chiral symmetry. The lagrangian in
Eq.~(\ref{eq:Lag1}) becomes 
gauge invariant if the ordinary
derivatives are replaced with the covariant derivatives
\begin{eqnarray}
\partial_{\mu} \xi      &  \rightarrow & \partial _{\mu}\xi 
             +ie A_{\mu}\Bigl[{\tau_3 \over 2},\xi\Bigr]
      \ , \label{eq:xider} \\
\partial_{\mu} N & \rightarrow & \Bigl(\partial _{\mu}
            +{i\over 2}eA_{\mu}(1+\tau_3)\Bigr)N\ ,\\
\partial_{\mu} \rho_\nu    &  \rightarrow & \partial _{\mu}\rho_\nu
             +ie A_{\mu}\Bigl[{\tau_3 \over 2},\rho_\nu\Bigr]
      \ . \label{eq:rhoder}
\end{eqnarray}
As a result, the axial and vector pion fields become
\begin{eqnarray}
\widetilde{a}_{\mu}    & = &
        a_{\mu} +{1\over 2}eA_{\mu}\Big (
             \xi\,[{\tau_3\over 2},\xi^{\dagger}]
             -\xi^{\dagger}\,[{\tau_3\over 2},\xi ]\Big) \ , \\
\widetilde{v}_{\mu}      & = &
       v_{\mu} +{1\over 2}eA_{\mu}\Big (
             \xi\,[{\tau_3\over 2},\xi^{\dagger}]
             +\xi^{\dagger}\,[{\tau_3\over 2},\xi ]\Big) \ ,
\end{eqnarray}
respectively, and the pion ``kinetic'' term is
\begin{equation}
  \fpi^2{\rm tr}\, (\widetilde{a}_\mu\widetilde{a}^\mu)
       = \fpi^2 {\rm tr}\,(a_{\mu} a^{\mu})
                   -2e\fpi^2 A^{\mu}{\rm tr}\,(v_{\mu}\tau_3)
                     + \cdots \ .
\end{equation}
%

In addition to terms arising from the preceding minimal substitutions,
there can be other non-minimal terms in the general effective lagrangian. 
For example, it is possible to have a direct rho-photon coupling 
term \cite{KROLL67}
\begin{equation}
-{e \over 2g_{\gamma}}F_{\mu\nu}\,{\rm tr}\,
           (\tau_3 \widetilde{\rho}^{\mu\nu}) \ ,  
\end{equation}
where $\widetilde{\rho}^{\mu\nu}$ is given by Eq.~(\ref{eq:rhomunu})
with the replacements in Eqs.~(\ref{eq:xider}) and (\ref{eq:rhoder}),
and $F_{\mu\nu}$ is the usual electromagnetic field tensor. 
In principle, a similar term proportional to $F_{\mu\nu}\,{\rm tr}\,
(\tau_3 \widetilde{v}^{\mu\nu})$ is also possible, although VMD 
applied to the pion form 
factor suggests that the coefficient of this term is small \cite{BROWN86}. 
Furthermore, the photon can couple to the nucleon as
\begin{equation}
-{e \over 4M}F_{\mu\nu}
  \overline N \lambda   \sigma^{\mu\nu} N
    +{e\over 2M^2}\partial_\mu F^{\mu\nu}
          \overline{N}\gamma_{\nu}(\beta_{\rm s}+\beta_{\rm v}\tau_3)N
     + \cdots
                 \ , \label{newterms}
\end{equation}
where
\begin{equation}
\lambda \equiv {1\over 2} \lambda_{\rm p} (1+\tau_3)
           + {1\over 2} \lambda_{\rm n} (1-\tau_3)\ , \label{eq:lambda}
\end{equation}
with $\lambda_{\rm p}=1.793$ and $\lambda_{\rm n}=-1.913$
the anomalous magnetic moments of the proton and the neutron, respectively. 
The first term in Eq.~(\ref{newterms}) is needed to
reproduce the magnetic moments of the nucleons, and the second one
contributes to the rms charge radii.
Note that these terms have $\nu=3$ and $\nu=4$, respectively.

We may now write down the resulting manifestly gauge-invariant lagrangian, 
which contains the preceding ``tilde'' fields. 
Instead of showing this lagrangian in its entirety, however, we omit terms 
that are irrelevant to our applications here.
The remaining lagrangian is manifestly chiral invariant,
except for some terms involving the photon field that are
proportional to the electric charge.
We find
\begin{eqnarray}
{\cal L}_2(x) &=& {\cal L}_1(x) - {1\over 2}e A_\mu 
              \overline N \gamma^{\mu}(1+\tau_3)N
       -{e \over 2g_{\gamma}}F_{\mu\nu}\,{\rm tr}\,
           (\tau_3 \rho^{\mu\nu})
           -2e\fpi^2 A^{\mu}{\rm tr}\,(v_{\mu}\tau_3) 
                \nonumber \\
         & &
               -{e \over 4M}F_{\mu\nu}
  \overline N \lambda   \sigma^{\mu\nu} N
           +{e\over 2M^2}\partial_\mu F^{\mu\nu}
    \overline{N}\gamma_{\nu}(\beta_{\rm s}+\beta_{\rm v}\tau_3)N
    - {1\over 4}F_{\mu\nu}F^{\mu\nu}
                   \ . \label{eq:Lmp}
\end{eqnarray}

It is well known that the isoscalar-vector $\omega$ meson
is needed to describe the short-range repulsion of the NN interaction.  
To have a realistic description of nuclear systems, the 
$\omega$ meson should be included.
Note that once the $\omega$ meson is introduced, 
an isoscalar-vector four-fermion term $(\overline{N} \gamma_\mu N)^2$ can 
be completely accounted for by adjusting $\gomega$ to fit experimental data,
as discussed above for the rho meson.

With respect to chiral $SU(2)$ symmetry, the  $\omega$ meson
can be treated 
as a chiral singlet represented by a vector
field $\omegaV_\mu(x)$. 
The relevant lagrangian can be written as
\begin{eqnarray}
{\cal L}_\omegav(x) &=& -{1\over 4}\omegaV_{\mu\nu}\omegaV^{\mu\nu}
          +{1\over 2}\momega^2\omegaV_\mu\omegaV^\mu
         -\gomega \overline N \gamma^{\mu}\omegaV_\mu N
          \nonumber \\
        & &
           -{\fomega \gomega \over 4M}\overline N
          \omegaV_{\mu\nu}  \sigma^{\mu\nu} N
         -{e \over 2g_{\gamma}}{1\over 3} F_{\mu\nu}\omegaV^{\mu\nu}
              + \cdots  \ ,\label{eq:Lomega}
\end{eqnarray}
where $\omegaV_{\mu\nu} \equiv \partial_\mu \omegaV_\nu - \partial_\nu
\omegaV_\mu$.
The kinetic terms again reflect the spin-1 nature of the particle.
Analyses of NN scattering show that the $\omega$ tensor coupling is 
not important, since $\fomega$ is small,
and this is indeed also the case from our fits (see below).
The factor of $1/3$ in the coupling to the photon arises naturally from
$SU(3)$ symmetry and the assumption of 
ideal mixing of the vector $\phi$ and $\omega$ mesons. 
The ellipsis represents higher-order terms involving additional powers of 
the fields and their derivatives, some of which will be considered 
explicitly in Sec.~II.E. 

Before ending this section, we note that it is consistent to integrate
out the vector $\phi (1020)$ field because it has a small coupling to the
nucleon. 
The effects of the vector $\phi$ in the isoscalar-vector channel
can be absorbed into $\gomega$, as discussed 
near the end of Sec.~II.A. For example, if we integrate out the
$\phi$, we obtain a contribution to $\beta_{\rm s}$
of $-2\sqrt{2}M^2g_\phi/(3m_\phi^2g_\gamma)$ and a four-fermion term
$(\overline N\gamma_\mu N)^2$ with a coefficient proportional
to $g_\phi^2/m_\phi^2$, where $g_\phi$ is its coupling constant to
the nucleons and $m_\phi$ is its mass. The contribution of 
this latter term is absorbed
into the renormalization of $\gomega$.
In general, one need only keep the lowest-lying meson in a given
channel at low energy.

\subsection{The Isoscalar-Scalar Channel}

Unlike the vector channels, the absence of well-defined, low-lying 
resonances in the isoscalar-scalar channel makes
the dynamics more difficult to identify and to model.
It is well known from the empirical NN scattering 
amplitude\cite{MCNEIL83,SHEPARD83,CLARK83}, from 
one-boson-exchange models of the NN 
interaction\cite{MACHLEIDT}, and from fits to
the properties of finite nuclei \cite{HOROWITZ} that there is a strong, 
mid-range
NN attraction arising from the isoscalar-scalar channel, with a range
corresponding to a mass of roughly 500 MeV.
It has also been shown in calculations based on either phenomenological
scattering amplitudes \cite{JACKSON75,DURSO77,DURSO80}
or explicit chirally symmetric models \cite{LIN89,LIN90} that an
attraction of the appropriate strength can be generated dynamically by 
correlated two-pion exchange between nucleons.
No nearby underlying resonance at the relevant mass is 
needed.\footnote{We note, however, that there are some recent indications 
for a narrow, light scalar\protect\cite{SVEC92}.}
Nevertheless, this dynamics is efficiently, conveniently, and adequately
represented by the exchange of a light isoscalar-scalar degree of freedom.
Therefore, in analogy to what we did with the vector mesons, we will 
introduce an effective scalar meson with a Yukawa coupling to the nucleon
to incorporate the observed mid-range
NN attraction, and thereby avoid the explicit computation of 
multi-loop pion contributions.
Based on the results of the two-pion-exchange calculations and
the success of one-boson-exchange models, it is sufficient to give the
scalar a well-defined mass, which we will choose here by fitting to
properties of finite nuclei.
Moreover, the light scalar is a chiral singlet, so it does not spoil
the chiral symmetry of the model.


In addition to the evidence from the NN interaction, one can ask what
other information exists on the dynamics in the isoscalar-scalar
channel.
A natural place to look is the physics of broken scale invariance in QCD.
This leads to an anomaly in the trace of the QCD energy-momentum tensor and
also to ``low-energy theorems'' on connected Green's functions
constructed from this trace \cite{VAINSHTEIN82}.
In pure-glue QCD, these low-energy theorems can be satisfied by introducing
a scalar glueball field (``gluonium'') with a mass of roughly 1.6 GeV.
When light quarks are added and the chiral symmetry is dynamically broken,
the trace anomaly can acquire a contribution in addition to that from
gluonium.
In Ref.~\cite{vacuum}, we considered this additional contribution to arise 
from the self-interactions of the light scalar mentioned above, and used the
low-energy theorems to constrain these interactions.

The lagrangian, which contained the light scalar, heavy scalar (glueball),
and other degrees of freedom, was scale invariant, except for 
some minimal scale-dependent self-interactions among the glueballs and the
light scalar. 
Since the glueballs are heavy, they can be integrated out
along with other heavy particles and high-momentum modes.
If one assumes no mixing between the scalar and the glueball, the scalar
potential is constrained to satisfy the low-energy theorems (at tree level), 
and the potential is given by\cite{MIRANSKY,vacuum}
\begin{equation}
      V_{\rm S} =   {d^2\over 4}m_{\rm s}^2S_0^2\bigg ({S^2 \over S_{ 0}^2}
         \bigg)^{2 / d}    \bigg ( {1 \over 2d}
              \ln {S^2 \over S_{ 0}^2}
                   -{1 \over 4} \bigg ) \ .
\end{equation}
Here $S_0$ is the vacuum expectation value of the light scalar field $S$,
$m_{\rm s}$ is its mass, and $S$ is assumed to have 
scale dimension $d$; that is, when $x\rightarrow \lambda^{-1} x$,
$S(x)\rightarrow \lambda^d S(\lambda x)$.

If we introduce the fluctuation field $\phi (x) \equiv S_0 - S(x)$ and
expand $V_{\rm S}$ as a power series in $\phi$, the coefficients are all
determined by the three parameters $S_0$, $m_{\rm s}$, and $d$.
The results of Ref.~\cite{vacuum} showed, however, that finite-nucleus 
observables
are primarily sensitive only to the first three terms in this expansion;
thus, even in this extremely simple model of the scale breaking, the 
low-energy theorems do not place significant constraints on the 
self-interactions of the light scalar field.
If we allow for the possibility of scalar--glueball mixing and 
employ field redefinitions to simplify the interaction terms between 
the light scalar and the nucleons, the form of the potential $V_{\rm S}$
can be modified and additional parameters need to be introduced.
Therefore, to take advantage of a simple Yukawa coupling to
the nucleon, we adopt here a general potential for the scalar meson,
which can be expanded in a Taylor series:
\begin{equation}
      V_{\rm S} = m_{\rm s}^2\phi^2\bigg ({1\over 2}
              +{\kappa_3\over 3!}\,{g_{\rm s}\phi \over M}
             + {\kappa_4 \over 4!}\,{g_{\rm s}^2\phi^2\over M^2}
             +\cdots \bigg ) \ . \label{eq:SP}
\end{equation}
Here we have anticipated the discussion of naive dimensional analysis
in the next subsection by including a factor of $1/\fpi$ for each power of
$\phi$; these factors are then eliminated in favor of 
$g_{\rm s}\approx M/\fpi$. 
Various counting factors are also included, as discussed in the
next subsection.
If the naturalness assumption is valid, the parameters
$\kappa_3$, $\kappa_4$, \dots\ should be of order unity. 
This is indeed supported by our fits to finite nuclei, as shown 
in Sec.~V.
As with the vector mesons discussed earlier, 
only one low-mass scalar field is needed to a good approximation.

From a different point of view, chiral symmetry allows for 
fermion contact terms, such as $(\overline NN)^2$, $[\partial_\mu
(\overline NN)]^2$, $(\overline NN)^3$, and so on.
These terms are relevant if there are strong interactions between nucleons
in the isoscalar-scalar channel.
By introducing a scalar field with mass $m_{\rm s}$, Yukawa coupling
$g_{\rm s}$, and nonlinear couplings $\kappa_3$, $\kappa_4$, 
\dots\ adjusted to fit experimental data, we can account for all of the 
contact terms.
The only relevant questions are which implementation allows for the most
efficient truncation and how many terms are needed in practical
calculations.

We shall not include an isovector-scalar field in our effective lagrangian
since the NN interaction in that channel is weak \cite{MACHLEIDT}.
There is no meson  with these quantum numbers with a mass below 1~GeV, 
and two (identical)
pions in a $J=0$ state cannot have $T=1$.

\subsection{Naive Dimensional Analysis and Field Redefinitions}

The lagrangian constructed thus far is based on symmetry considerations
and the desire to reproduce realistic electromagnetic and NN interactions
while keeping pion-loop calculations to a minimum.
We have illustrated the lowest-order terms in powers of the fields and
their derivatives, but in principle, an infinite number of terms is possible,
and we must have a meaningful way to truncate the effective lagrangian for
the theory to have any predictive power.

A naive dimensional analysis (NDA) for assigning a coefficient of the 
appropriate size to any term in an effective lagrangian 
has been proposed
by Georgi and Manohar\cite{GEORGI84b,GEORGI93}. 
The basic assumption of ``naturalness'' is that once the appropriate
dimensional scales have been extracted using NDA, 
the remaining overall dimensionless
coefficients should all be of order unity.
For the strong interaction, there are two relevant scales:
the pion-decay constant $\fpi \approx 93~\mbox{MeV}$ 
and a larger scale $0.5 \alt \Lambda \alt 1~\mbox{GeV}$, 
which characterizes the mass scale of physics beyond Goldstone bosons.
The NDA rules for a given term in the lagrangian density are
\begin{description}
\item[1.] Include a factor of $1/\fpi$ for each strongly interacting
        field.
\item[2.] Assign an overall factor of $\fpi^2 \Lambda^2$.
\item[3.] Multiply by factors of $1/\Lambda$ to achieve dimension (mass)$^4$. 
\end{description}
The appropriate mass for $\Lambda$
may be the nucleon mass $M$ or a non-Goldstone
boson mass; the difference is not important for verifying naturalness
but can be relevant for $N_c$ counting arguments \cite{FSTprepare}.

As noted by Georgi\cite{GEORGI93}, Rule~1 states that the 
amplitude for generating any 
strongly interacting particle is proportional to the amplitude $\fpi$
for emitting a Goldstone boson.
Rule~2 can be understood as an
overall normalization factor that arises from the standard way
of writing the mass terms of non-Goldstone bosons
(or, in general, the quadratic terms in the lagrangian). 
For example, one may
write the mass term of an isoscalar-scalar field $\phi(x)$ as
\begin{equation}
     {1\over 2}m^2_{\rm s} \phi^2 =
        {1\over 2}\fpi^2 \Lambda^2 
         {m_{\rm s}^2 \over \Lambda^2}
         {\phi^2 \over \fpi^2} \ ,
\end{equation}
where the scalar mass $m_{\rm s}$ is treated as roughly the same
size as $\Lambda$.
By applying Rule~1 and extracting the overall factor of $\fpi^2 \Lambda^2$,
the dimensionless coefficient is  of $O(1)$.
Terms with gradients (or external fields) will be associated with one power of
$1/\Lambda$ for each derivative (or field) as a result of integrating
out physics above the scale $\Lambda$.
(A simple example is the expansion at low momentum 
of a tree-level propagator for a heavy 
meson of mass $m_H$, which leads to terms with powers of $\partial^2/m_H^2$.)
It is because of these $1/\Lambda$ suppression factors and dimensional 
analysis that one arrives at Rule~3.
 
After the dimensional NDA factors 
and some appropriate counting factors are extracted (such as $1/n!$ for the 
$n^{\rm th}$ power $\phi^n$; see below), 
the naturalness assumption implies that any overall dimensionless 
coefficients should be of order unity.
Without such an assumption,  an effective lagrangian will
not be 
predictive.\footnote{The assumption of renormalizability would also lead to
a finite number of parameters and well-defined predictions, but we will
not consider this option here\protect\cite{SEROT86,SEROT92}.}
Until one can derive the effective lagrangian from QCD, the naturalness 
assumption must be checked by fitting to experimental data.

As an example of NDA and naturalness, 
we can estimate the sizes of the coupling
constants $g_{\rm  \scriptscriptstyle A}$, $g_{\rm s}$,
and $\gomega$ by applying the rules to the following interaction terms: 
$g_{\rm \scriptscriptstyle A} \overline N \gamma^{\mu}\gamma_5a_{\mu}N$, 
$g_{\rm s}\overline N \phi N$, and 
$\gomega \overline N\gamma^{\mu}\omegaV_\mu N$.
Note that the pion field in $a_{\mu}$ is already associated with
an overall factor of $1/\fpi$ [see Eq.~(\ref{eq:leading})] 
so $a_\mu$ is simply counted as a derivative. 
It is straightforward to find that
\begin{equation}
g_{\rm \scriptscriptstyle A}\sim 1 \ , \ \  \ \ 
g_{\rm s},\ \gomega \sim {\Lambda\over \fpi}\alt 4\pi \ . \label{eq:ggg}
\end{equation}
That is, Eq.~(\ref{eq:ggg}) gives the natural sizes of these 
couplings,
where we use the symbol $\sim$ to denote order of magnitude.%
\footnote{The bound $\Lambda \alt 4\pi \fpi$ was
originally proposed by Georgi and Manohar \cite{GEORGI84b}
and subsequently refined to $\Lambda \alt 4\pi \fpi/\sqrt{N_f}$, where $N_f$
is the number of light quark flavors \cite{GOLDEN}.
QCD appears to essentially saturate this bound.}
It is reassuring that the preceding values are indeed consistent with 
phenomenology.

According to NDA,
a generic term in the effective lagrangian involving the isoscalar
fields and the nucleon field can be written as
(generalizing 
to include the pion, rho, and photon is straightforward)
\begin{equation}
  g {1 \over m!}{1\over n!} \fpi^2 \Lambda^2 
    \biggl( {\overline NN\over \fpi^2\Lambda} \biggr)^l
    \biggl(  {\phi\over \fpi} \biggr)^m
    \biggl(  {V\over \fpi} \biggr)^n
    \biggl( {\partial\over \Lambda} \biggr)^p 
    \ , \label{eq:generic}  
\end{equation}
which includes $l$ nucleon bilinears, $m$ scalar fields, $n$ vector fields,
and $p$ derivatives.
Gamma matrices and Lorentz indices have been suppressed.
The coupling constant $g$ is dimensionless [and of $O(1)$ if naturalness
holds].


Numerical counting factors for terms with multiple powers of 
meson fields 
are included because the NDA rules are actually meant to apply to the 
tree-level amplitude generated by the corresponding vertex.
Thus, 
a term containing $\phi^m$, for example, should have a factor $1/m!$
associated with it.
These additional factors are certainly relevant when, say, $m=4$. 
An explicit illustration of the origin of the counting factors can be
found in the pion mass term in ChPT:
\begin{equation}
 {1\over 4}m_\pi^2\fpi^2 {\rm tr}\, (U+U^\dagger-2) =
   m_\pi^2\fpi^2\sum_{n=1}^{\infty} (-1)^n
        {(\bbox{\pi}^2)^n \over (2n)! \fpi^{2n} } \ . 
\end{equation}

An important feature of the effective lagrangian is that  
each derivative on a field is associated with a factor of $1/\Lambda$. 
Thus organizing the lagrangian in powers of
derivatives can be a good expansion for describing processes where the
dominant energies and momenta are not too large. 
Nevertheless, this organization is intimately connected with the important
issue of how to remove redundant terms in the effective lagrangian.
By employing the freedom to redefine the fields,
one can organize the lagrangian such that there are no terms with 
$\partial^2$ acting on a boson field or $\gamma^\mu\partial_\mu$ 
acting on a fermion 
field, {\it except} in the kinetic-energy terms.
These field redefinitions do not change the on-shell $S$-matrix 
elements\cite{GEORGI91,COLEMAN69,CALLAN69,BANDO88}.

In finite-density applications, however,
the non-Goldstone bosons exchanged by the 
nucleons are far off shell,
which motivates a different definition of the corresponding field
variables.
Short-distance effects due to heavier particles, high-momentum modes, 
and dynamical vacuum contributions
are assumed to be integrated out, so that
their consequences appear only indirectly in the coefficients of the 
lagrangian.
Derivatives on the non-Goldstone
fields are already small 
compared to their masses in our applications because the momenta are
limited by the Fermi momentum. 
Thus, we do not need to eliminate these derivatives.
 
Instead, we can use nonlinear field redefinitions to simplify the form of 
the interaction terms between the nucleons and the non-Goldstone bosons
so that they take  the conventional Yukawa form.
This ensures the consistency of the power counting rule (\ref{eq:order}). 
Recall that one of the main purposes for introducing the non-Goldstone 
bosons is to describe the intermediate-range NN interactions. 
Since the nonlinear field redefinitions do not change the quadratic kinetic
and mass terms, no changes occur to the ranges of the 
one-boson-exchange interactions in 
free space. 
All other heavy boson-exchange contributions are short-ranged, so we
can efficiently parametrize them using meson self-couplings.
We can easily show the invariance of the one-loop energy functional under 
field redefinitions, although a general proof to all orders seems difficult. 
It is still convenient to use the field redefinitions to 
remove all derivatives on the nucleon fields except in the kinetic term.
These derivatives can be large because valence-nucleon
energies are of the order of the nucleon mass $M$.

At finite density, the Fermi momentum appears as a new scale 
and meson fields develop expectation values.
We must then also decide how to organize the nonderivative terms.
The key observation is that at normal nuclear-matter density 
(or in the center of a heavy nucleus),
the scalar and vector mean-fields are typically 
\begin{equation}
0.25 \, M\ \lesssim\ g_{\rm s}\phi_0,\  \gomega V_0\ \lesssim\ 0.4 \, M\ ,
   \label{eq:fdest}
\end{equation}
and the gradients of the mean fields in the nuclear surface, where they 
are largest, are 
\begin{equation}
g_{\rm s}|\bbox{\nabla}\phi_0|,\  
\gomega |\bbox{\nabla}V_0 | \approx  (0.2 M)^2  \ .
                       \label{eq:derest}
\end{equation}
(Note that the {\em magnitudes\/} of the mean fields are reduced in 
the surface.)
  Equations~(\ref{eq:fdest}) and (\ref{eq:derest}) are our basis
for estimating the leading contributions to the energy
of any term in our lagrangian.  

To summarize, the assumption of naturalness and the observation that
the mean fields and their derivatives are small compared to $M$ up
to moderate densities allows us to organize the lagrangian in powers of the 
fields and their derivatives.
We will take advantage of the freedom to redefine the fields to simplify 
the interaction terms between the nucleons and the non-Goldstone bosons. 
For the nucleons, pions, and photons, we use field redefinitions to remove 
high-order derivatives, whereas the scalar and vector meson parts 
of the lagrangian can contain powers of $\partial^2$ in addition to the
kinetic terms, since these terms are small.
By truncating the lagrangian at a given order in $\nu$, 
we have a finite number of 
parameters to determine through fitting to nuclear observables.

\subsection{The Full Lagrangian}

Based on the preceding discussion, we illustrate here how nonlinear
field redefinitions allow us to simplify the terms involving interactions 
between the non-Goldstone bosons and the nucleons. 
Derivatives on the nucleon
field other than the kinetic term are  removed by redefining the field
because these derivatives on the nearly on-shell nucleons are
of $O(M)$.  
For the photons and the Goldstone pions,  field redefinitions are used 
to remove higher derivatives \cite{GEORGI91}, 
so that their interactions with the nucleons 
are unconstrained except by gauge invariance and 
chiral symmetry.
Through $\nu=4$,
the part of the effective lagrangian involving nucleons can then be 
written as
\begin{eqnarray}
{\cal L}_{\rm N}(x) &=&
         \overline N \Big (i\gamma^{\mu}{\cal D}_\mu
             +g_{\rm \scriptscriptstyle A}\gamma^{\mu}\gamma_5a_{\mu}
              -M +g_{\rm s }\phi \Big )N
                    \nonumber \\[3pt]
         & &
       -{f_\rho g_\rho \over 4M}\overline N
        \rho_{\mu\nu}  \sigma^{\mu\nu} N
         -{\fomega \gomega \over 4M}\overline N\omega_{\mu\nu}
                  \sigma^{\mu\nu} N
           - {\kappa_\pi\over M} \overline{N}
              v_{\mu\nu}\sigma^{\mu\nu}  N
                     \nonumber \\[3pt]
       & & -{e \over 4M}F_{\mu\nu}
         \overline N \lambda   \sigma^{\mu\nu} N
           -{e\over 2M^2}
           \overline{N}\gamma_{\mu}(\beta_{\rm s}
               +\beta_{\rm v}\tau_3)N
               \partial_\nu F^{\mu\nu}
           \ , \label{eq:lnucl}
\end{eqnarray}
where the covariant derivative is 
\begin{equation}
   {\cal D}_\mu=\partial_{\mu}
            +iv_{\mu}+ig_{\rho}\rho_\mu
             +i\gomega\omegaV_\mu
             +{i\over 2}eA_\mu(1+\tau_3)       \ .
\end{equation}

One might expect more general 
isoscalar-scalar couplings to the nucleons.
However, all possible isoscalar-scalar field combinations
of the non-Goldstone boson fields that can couple to $\overline{N}N$
can be redefined in terms of a single isoscalar-scalar field
with a Yukawa coupling $\overline{N}N\phi$.
For example, the following couplings are redundant:
\[
             \overline N  N\phi^2\, ,\ \ \overline N  N\phi^3
           \, ,\ \ \ \overline N  N\partial^2\phi
            \, ,\ \ \ \overline N  N\omegaV_\mu\omegaV^\mu \ .
\]
The redefinition of the field induces changes in the coefficients of
the mesonic part of the lagrangian, but since all terms to 
a given order in $\nu$
are already included, there is no loss of generality. 
A similar observation holds for the isoscalar-vector and the isovector-vector 
couplings.

A term of the form  $\overline N \gamma^\mu N\partial_\mu\phi$
is irrelevant because the baryon current is conserved.
Moreover, a coupling of the form
\begin{equation}
\overline N  N  \partial_\mu \omegaV^\mu \label{eq:dV}
\end{equation}
can be removed from the hamiltonian
by using the constraint equation for the vector field.
 The identity
\begin{equation}
{\partial}_\mu\equiv{1\over 2}
    ( \gamma_\mu\gamma_\nu{\partial}^\nu
          +\gamma_\nu{\partial}^\nu\gamma_\mu )
        \label{eq:relation}
\end{equation}
and the equations of motion allow us to reduce a term of the form
$\overline{N}i\omegaV^\mu{\partial}_\mu N$ in the lagrangian to a sum of
Yukawa and tensor terms 
and the term in (\ref{eq:dV}) in the hamiltonian.
As noted, bilinears in the nucleon fields and derivatives acting on such 
bilinears can be eliminated in favor of low-mass non-Goldstone bosons.
Furthermore, we can use Eq.~(\ref{eq:relation}) and partial integrations 
to reduce mixed-derivative terms such as
$(\overline N\partial_\mu N)(\overline N\partial^\mu N)$ to the
form $\partial_\nu(\overline N\gamma^\sigma \gamma_\mu N) \partial_\sigma 
(\overline N
\gamma^\nu\gamma^\mu N)$, which we assume is small compared to terms that are
retained based on the minor role of tensor mesons in meson-exchange
phenomenology.

    In Eq.~(\ref{eq:lnucl}), we have omitted the following
$\nu=4$ terms:
\begin{equation}
 \overline N \rho_{\mu\nu}
                  \sigma^{\mu\nu} N\phi\, ,\ \
 \overline N\omegaV_{\mu\nu}
                   \sigma^{\mu\nu} N\phi \ . \label{eq:p5terms}
\end{equation}
The leading effect of the terms in (\ref{eq:p5terms}) is a slight
modification of the tensor couplings $f_\rho$ and $\fomega$ at finite
density due to the nonvanishing expectation value of the scalar field.
However, the effects of the tensor terms are themselves already small,
so the  further modification is negligible. 
Thus, while we may keep terms through $\nu=4$ in the lagrangian,
not all $\nu=4$ terms contribute equally in a given application.
In fact, the pion terms do not contribute to the energy of
nuclei at tree-level.

The mesonic part of the lagrangian 
is also organized in powers of the
fields and their derivatives.
Keeping terms through $\nu=4$, we find
\begin{eqnarray}
{\cal L}_{\rm M}(x) &=&
           {1\over 2}\Big(
                 1  + \alpha_1 {g_{\rm s}\phi\over M}
                \Big) \partial_\mu \phi \partial^\mu \phi
                   +{f_{\pi}^2\over 4}\, {\rm tr}\,
                  (\partial _{\mu}U\partial ^{\mu}U^{\dagger})
                \nonumber  \\[3pt]
       &  &  \null     
       -{1\over 2}{\rm tr}\, (\rho_{\mu\nu}\rho^{\mu\nu})
                   -{1 \over 4}\Big(
                 1  + \alpha_2 {g_{\rm s}\phi\over M}
                \Big) \omegaV_{\mu\nu}\omegaV^{\mu\nu}
           -g_{\rho \pi\pi}{2\fpi^2 \over m_\rho^2}
              {\rm tr}\,(\rho_{\mu\nu}v^{\mu\nu})
                          \nonumber  \\[3pt]
       &  &  \null     
              -2ef_{\pi}^2A^{\mu}{\rm tr}\,(v_{\mu}\tau_3)
              -{e \over 2g_{\gamma}}F_{\mu\nu}\Big[\,
           {\rm tr}\,(\tau_3\rho^{\mu\nu})
            +{1\over 3}\,\omegaV^{\mu\nu} \Big]  \nonumber  \\[3pt]
        &  &  \null  + {1\over 2}\bigg (
      1+ \eta_1 {g_{\rm s}\phi\over M}  
        + {\eta_2\over 2} {g_{\rm s}^2\phi^2\over M^2} \bigg )
                   \momega^2 \omegaV_{\mu}\omegaV^{\mu}
          +{1\over 4!}\zeta_0
             \gomega^2 (\omegaV_{\mu}\omegaV^{\mu})^2
       \nonumber \\[3pt]
         &  &       \null    +
            \bigg ( 1+ \eta_\rho {g_{\rm s}\phi\over M} \bigg )
             m_\rho^2 {\rm tr}\,(\rho_{\mu}\rho^\mu)
           - m_{\rm s}^2\phi^2\bigg ({1\over 2}+{\kappa_3\over 3!}\,
            {g_{\rm s}\phi \over M} + {\kappa_4 \over 4!}\,
           {g_{\rm s}^2\phi^2\over  M^2}\bigg )
                \ ,\label{eq:NLag}
\end{eqnarray}
where, apart from conventional definitions of some couplings
($g_s$, $\gomega$, and $g_\gamma$) and the masses, we
have defined the parameters so that they are of
order unity according to naive dimensional analysis. 
(Again, nuclei will test this hypothesis for us!)
Also, since the expectation value of the $\rho$ field is typically an order 
of magnitude smaller than that of the $\omegaV$ field, we have retained
nonlinear $\rho$ couplings only through $\nu=3$.
Note that the $\alpha_1$ and
$\alpha_2$ terms have $\nu=5$. However, from  
Eqs.~(\ref{eq:fdest}) and (\ref{eq:derest}) we can estimate that
these two $\nu=5$ terms are  numerically of the
same magnitude as the quartic scalar term in the nuclear surface energy, so
we have retained them. Thus, numerical factors such as
$1/ n!$, which are cancelled in scattering amplitudes,
 are  relevant in deciding the importance of a term in the energy.

Our full lagrangian through $\nu=4$ is then
\begin{equation}
   {\cal L}= {\cal L}_{\rm N}+{\cal L}_{\rm M}\ . \label{eq:fullL}
\end{equation}
For applications to normal nuclei, keeping these terms should be
sufficient if the parameters are natural. This may be 
checked by including  the following $\nu=5$ terms:
\begin{equation}
   {\cal L}_5 = 
                       - {1\over 5!}\kappa_5
                {g_{\rm s}^3\phi^3\over M^3} m_{\rm s}^2\phi^2
                         + {1\over 3!} \eta_3 
                {g_{\rm s}^3\phi^3\over M^3}
               \cdot {1\over 2}\, \momega^2 \omegaV_{\mu}\omegaV^{\mu}
                         + {1\over 4!}\zeta_1
                   {g_{\rm s}\phi\over M}
             \gomega^2 (\omegaV_{\mu}\omegaV^{\mu})^2 \ .
                      \label{eq:fifth}
\end{equation}
As discussed below, the inclusion of ${\cal L}_5$ 
improves our fits only marginally.

\section{Electromagnetic Structure of the pion and nucleon}

For an effective lagrangian to be useful for calculations 
of nuclear matter and finite nuclei, it should incorporate
the low-energy structure of the pion and the nucleon.
This physics has been extensively studied;
there is a long history of vector-meson dominance (VMD) descriptions
of electromagnetic form factors \cite{VMDa,VMDb,VMDc,BROWN86}.
We illustrate here how 
the electromagnetic form factors of the pion
and the nucleon at low momenta arise from our effective
lagrangian through VMD and also through
direct couplings of photons to pion and
nucleon fields \cite{IACHELLO73}.

The electromagnetic (EM) current can be obtained from the 
lagrangian (\ref{eq:lnucl}), (\ref{eq:NLag}), and (\ref{eq:fullL})
by taking $\delta{\cal L} /\delta (eA_\mu)$ after some partial integration:
\begin{eqnarray}
 J^\mu &=& {1\over 2}\overline{N}(1+\tau_3)\gamma^\mu N
       +{1\over 2M}\partial_\nu(\overline{N}\lambda\sigma^{\mu\nu}N)
         -{1\over 2M^2}\partial^2[\overline{N}
           (\beta_{\rm s}+\beta_{\rm v}\tau_3)\gamma^\mu N] 
          \nonumber  \\[4pt]
     & &    +  {1\over g_\gamma}(\partial_\nu \rho_3^{\mu\nu}
             +{1\over 3} \partial_\nu\omegaV^{\mu\nu})
         +2 \fpi^2{\rm tr}(v^\mu\tau_3) \ . \label{eq:EMcrnt}
\end{eqnarray}
Note that the photon can couple to two or more
pions either directly or
through the exchange of a neutral rho meson at non-zero momentum
transfers. The coupling of the photon to the
nucleons is also direct or through the exchange of neutral vector
mesons (rho or omega). 

We can determine the tree-level EM form factor of the pion $F(-q^2)$  
from the current (\ref{eq:EMcrnt}) and
the lagrangian (\ref{eq:NLag}). 
For spacelike momentum transfers $Q^2=-q^2$, we have
\begin{equation}
F(Q^2)=1-{g_{\rho\pi\pi}\over g_\gamma}\,
       {Q^2 \over Q^2+m_\rho^2}+\cdots \ . \label{eq:piF}
\end{equation}
Thus the mean-square charge radius of the pion is
\begin{equation}
\langle r^2_\pi \rangle \equiv -6{{\rm d}F(Q^2) \over {\rm d} Q^2}
        \Bigg|_{Q^2=0}
       = {6g_{\rho\pi\pi}\over g_\gamma m_\rho^2}=(0.69\,{\rm fm})^2 \ ,
\end{equation}
where we have used the experimental values\cite{SEROT86,HAKIOGLU91}
\begin{equation}
{g_{\rho\pi\pi}^2\over 4\pi}=2.9 \, , \ \ \ \ 
    {g_\gamma^2\over 4\pi}=2.0  \ ,  \label{eq:ggamma}
\end{equation}
which reproduce the rho widths 
$\Gamma_{\rho\rightarrow \pi\pi}=150\,$MeV
and
$\Gamma_{\rho^0\rightarrow e^+e^-}=6.8\,$keV, respectively.
The observed charge radius is
$\langle r^2_\pi \rangle^{1/2}_{\rm exp}=(0.66\pm 0.01)\, {\rm fm}$.

Similarly, we can determine the  EM isoscalar and isovector charge form
factors of the nucleon,
\begin{eqnarray}
F_1^{\rm s}(Q^2) &=& {1\over 2} - {\beta_{\rm s}\over 2}\,{Q^2 \over M^2}
     -{\gomega\over 3g_\gamma}\,
       {Q^2 \over Q^2+\momega^2}+\cdots \ ,     \\[3pt]
F_1^{\rm v}(Q^2) &=& {1\over 2} - {\beta_{\rm v}\over 2}\,{Q^2 \over M^2}
     -{g_\rho\over 2g_\gamma}\,
       {Q^2 \over Q^2+m_\rho^2}+\cdots  \ ,
\end{eqnarray}
and the anomalous form factors
\begin{eqnarray}
F_2^{\rm s}(Q^2) &=& {\lambda_p+\lambda_n\over 2}
     -{\fomega \gomega\over 3g_\gamma}\,
       {Q^2 \over Q^2+\momega^2}+\cdots \ ,   \\[3pt]
F_2^{\rm v}(Q^2) &=& {\lambda_p-\lambda_n\over 2} 
     -{f_\rho g_\rho\over 2g_\gamma}\,
       {Q^2 \over Q^2+m_\rho^2}+\cdots \ .
\end{eqnarray}
The corresponding mean-square charge radii are
\begin{eqnarray}
\langle r^2 \rangle_{{\rm s}1} &=& 
      6\biggl( {\beta_{\rm s}\over M^2}
                 +{2\gomega\over 3g_\gamma \momega^2}
                 \biggr)  \ ,\label{eq:size1} \\[3pt]
\langle r^2 \rangle_{{\rm v}1} &=&
      6\biggl( {\beta_{\rm v}\over M^2}
                 +{g_\rho\over g_\gamma m_\rho^2}\biggr)\ ,
\end{eqnarray}
and
\begin{eqnarray}
\langle r^2 \rangle_{{\rm s}2} &=&
      {4\over \lambda_p+\lambda_n}\,
      {\fomega \gomega\over g_\gamma \momega^2} \ , \label{eq:sizze}\\[3pt]
 \langle r^2 \rangle_{{\rm v}2} &=&
      {6\over \lambda_p-\lambda_n}\,
      {f_\rho g_\rho\over g_\gamma m_\rho^2} \ . \label{eq:size2}
\end{eqnarray}
We will fit $\gomega$ and $g_\rho$ to the properties of finite nuclei
and then choose $\beta_{\rm s}$, $\beta_{\rm v}$, and $f_\rho$ 
to reproduce the 
experimental charge radii of the nucleon\cite{BROWN86}:
\begin{eqnarray}
\langle r^2 \rangle_{{\rm s}1}^{1/2} &\approx&
\langle r^2 \rangle_{{\rm v}1}^{1/2} \approx 
      0.79 \, {\rm fm}  \ , \label{eq:nucsize1}\\
\langle r^2 \rangle_{{\rm v}2}^{1/2}
        &=&(0.88\pm 0.05)\, {\rm fm}\ . \label{eq:nucsize2}
\end{eqnarray}
  We do not use Eq.~(\ref{eq:sizze}) 
  to fix $f_{\rm v}$ because the remaining radius
$\langle r^2 \rangle_{{\rm s}2}^{1/2}$ is not
sufficiently well determined \cite{BROWN86}.

  As in Refs.~\cite{BROWN86} and \cite{IACHELLO73},
we have a contribution
from vector dominance and a correction to the intrinic structure
to second order in a derivative expansion, that is,
to order $Q^2$.  This correction is adequate for our applications to
nuclear structure; 
applications involving higher
momenta would require us to include additional
nonrenormalizable interaction terms \cite{LEPAGE89}
and loop effects.
Thus we do not expect our nucleon form factors to be accurate at large
momentum transfers.

\section{Finite Nuclei and Nuclear Matter}

We work at one-baryon-loop order in this paper, which is
equivalent to the Dirac--Hartree approximation \cite{HOROWITZ}. 
This should be sufficient to test the consistency of 
NDA and the truncation of our lagrangian.
As discussed in Ref.~\cite{bodmer}, a Hartree calculation can be viewed
as equivalent to a density-functional approach, in which higher-order
many-body corrections are treated approximately.
Based on Dirac--Brueckner--Hartree--Fock calculations of nuclear matter,
exchange and correlation corrections to the nucleon self-energies are
expected to be small \cite{HOROWITZ87}.
The stability of the Hartree results will be tested in a two-loop 
calculation, which will be presented elsewhere.

The derivation of the Hartree equations and energy functional for nuclei
is similar to that described in Ref.~\cite{vacuum}. 
Because of the large masses of the nucleons and non-Goldstone bosons,
loop integrals involving these particles include dynamics from distance
scales that are much shorter than the scale set by the valence-nucleon
momenta, which are limited by the Fermi momentum $k_{\rm F}$.
As discussed above, these short-range effects are included implicitly in the 
coefficients of the lagrangian. 
Formally, one can include counterterms to remove these loop effects to all 
orders, which is always possible, since {\it all such terms are already 
contained in the effective lagrangian}.

Our present approach differs from Ref.~\cite{vacuum} in that we take
a more general point of view and employ field redefinitions to
simplify the couplings between the nucleon and the non-Goldstone bosons.
The structure of the nucleon is now described {\it within the theory\/}, 
so that no external electromagnetic form factors are needed, as in earlier
calculations \cite{HOROWITZ,REINHARD89,LosAlamos}. 
By taking $\kappa_3$, $\kappa_4$, etc. in the scalar potential  
as free parameters, we also allow for a more general description of the 
dynamics in the isoscalar-scalar channel, as well as for modifications 
from field redefinitions.

The single-particle Dirac
hamiltonian can be written as \cite{vacuum}
\begin{eqnarray}
  h({\bf x}) &=& -i\bbox{\alpha\cdot\nabla}
             +W({\bf x})
            +{\textstyle{1\over 2}}\tau_3 R({\bf x})
          + \beta \bigl( M-\Phi({\bf x}) \bigr)
          +{\textstyle{1\over 2}}(1+\tau_3)A({\bf x})
                \nonumber \\[4pt]
    & &    -{i\over 2M}\beta \bbox{\alpha\,\cdot\,}
        (f_\rho{\textstyle{1\over 2}}\tau_3\bbox{\nabla}R 
          + \fomega\bbox{\nabla}W)
      +{1\over 2M^2} (\beta_{\rm s} + \beta_{\rm v}\tau_3)
            \bbox{\nabla}^2A
      -{i\over 2M}\lambda\beta \bbox{\alpha\,\cdot\,}
                \bbox{\nabla}A \ ,
\end{eqnarray}
where $\beta = \gamma_0$, $\bbox{\alpha}=\gamma_0\bbox{\gamma}$,
and we have defined scaled mean fields by including the 
couplings \cite{BODMER91}:
%
\begin{eqnarray}
       W({\bf x}) &=& \gomega V_0({\bf x})\ , \label{eq:W} \\[4pt]
       \Phi({\bf x}) &=& g_{\rm s}\phi_0({\bf x})\ , \label{eq:Phi} \\[4pt]
       R({\bf x}) &=& g_\rho b_0({\bf x})\ , \label{eq:R} \\[4pt]
       A({\bf x}) &=& e A_0({\bf x})\ .  \label{eq:A}
\end{eqnarray}
%
The Dirac equation with eigenvalues
$E_\alpha$ and eigenfunctions $\psi_{\alpha}({\bf x})$
is\cite{HOROWITZ,SEROT86}
\begin{equation}
 h \psi_{\alpha}({\bf x}) = E_{\alpha} \psi_{\alpha}({\bf x})\ , \
\ \ \ \
     \int {\rm d}^3x\,
\psi^{\dagger}_{\alpha}({\bf x})\psi_{\alpha}({\bf x})
     = 1 \ .  \label{eq:norm}
\end{equation}

We follow the notation and conventions of Ref.~\cite{SEROT86} and
write the eigenfunctions for spherically symmetric nuclei as
\begin{equation}
  \psi_\alpha ({\bf x}) = \psi_{n\kappa mt} ({\bf x}) =
    \left( 
      \begin{array}{c}
          {\textstyle i\over\textstyle r}G_a(r) \Phi_{\kappa m}  \\ 
          -{\textstyle 1\over \textstyle r}F_a(r) \Phi_{-\kappa m}  
      \end{array}
     \right)
     \zeta_t 
     \ , \qquad a\equiv \{n,\kappa,m\}  \ , 
    \label{eq:fandg}
\end{equation}
where $\Phi_{\kappa m}$ is a spin spherical harmonic,
$t=1/2$ for protons, and $t=-1/2$ for neutrons.
The radial equations for $G$ and $F$ become
\begin{eqnarray}
  \biggl({d\over dr} + {\kappa\over r}\biggr) G_a(r) -
    [E_a - U_1(r) + U_2(r)] F_a(r) - U_3 G_a(r) &=& 0 \ , \\
  \biggl({d\over dr} - {\kappa\over r}\biggr) F_a(r) +
    [E_a - U_1(r) - U_2(r)] G_a(r) + U_3 F_a(r) &=& 0 \ ,
\end{eqnarray}
where we have defined the single-particle potentials
\begin{eqnarray}
  U_1(r) &\equiv& W(r) + t_a R(r) + (t_a+{1\over 2})A(r)
            + {1\over 2M^2}(\beta_{\rm s} +2t_a \beta_{\rm v}) 
            \nabla^2 A(r) \ , \\
  U_2(r) &\equiv& M - \Phi(r) \ ,\\ 
  U_3(r) &\equiv& {1\over 2M}\Bigl\{\fomega W'(r) + t_a f_\rho R'(r) + 
         A'(r)[ (\lambda_p+\lambda_n)/2 + t_a(\lambda_p-\lambda_n)]
         \Bigr\}\ ,
\end{eqnarray} 
and a prime denotes a radial derivative.

The mean-field equations for $\Phi$, $W$, $R$, and $A$ are
\begin{eqnarray}
   -{\bbox{\nabla}}^2 \Phi + m_{\rm s}^2 \Phi  & = &
     g_{\rm s}^2 \rho_{\rm s}({\bf x})
        -{m_{\rm s}^2\over M}\Phi^2
         \Big({\kappa_3\over 2}+{\kappa_4\over 3!}{\Phi\over M}
               +{\kappa_5\over 4!}{\Phi^2\over M^2}\Big)
                \nonumber  \\[4pt]
     & & \null      
       +{g_{\rm s}^2 \over 2M}
         \Big(\eta_1+\eta_2{\Phi\over M}
               +{\eta_3\over 2}{\Phi^2\over M^2}\Big)
                { \momega^2\over \gomega^2} W^2
              +{\zeta_1 g_{\rm s}^2\over 4! \gomega^2}{W^4\over M}
          \nonumber  \\[4pt]
     & & \null
             +{g_{\rm s}^2 \eta_\rho \over 2M}
            { m_\rho^2 \over g_\rho^2} R^2
       +{\alpha_1\over 2M}[
             (\bbox{\nabla}\Phi)^2+2\Phi{\bbox{\nabla}}^2 \Phi ]
        + {\alpha_2g_{\rm s}^2\over 2M\gomega^2}(\bbox{\nabla}W)^2
                                      \ , \\[4pt]
   -{\bbox{\nabla}}^2 W + \momega^2 W  & = &
        \gomega^2 [\rho_{\rm B}({\bf x})
                      +{\fomega\over 2M}\bbox{\nabla}\cdot
                      (\rho_{\rm B}^{\rm T}({\bf x})\widehat r) ]
        -\Big(\eta_1+{\eta_2\over 2}{\Phi\over M}
               +{\eta_3\over 3!}{\Phi^2\over M^2}\Big){\Phi\over M}
                 \momega^2 W        \nonumber  \\[4pt]
         & & \null
      -{1\over 3!}\Big( \zeta_0 +\zeta_1{\Phi\over M}\Big)W^3
          +{\alpha_2 \over M} (\bbox{\nabla}\Phi\bbox{\cdot\nabla}W
                    +\Phi\bbox{\nabla}^2W)
          -{e^2\gomega \over 3 g_\gamma}\rho_{\rm chg}({\bf x})
                                \ , \\[4pt]
       -{\bbox{\nabla}}^2 R  + m_\rho^2 R &=&
               {1\over 2}g_\rho^2[\rho_3({\bf x})
                      +{f_\rho\over 2M}\bbox{\nabla}\cdot
                      (\rho_3^{\rm T}({\bf x})\widehat r) ]
                  -\eta_\rho {\Phi\over M}
                 m_\rho^2 R -{e^2g_\rho \over g_\gamma}
                  \rho_{\rm chg}({\bf x})            \ , \\[4pt]
      -{\bbox{\nabla}}^2 A &=& e^2 \rho_{\rm chg}({\bf x}) \ .
\end{eqnarray}
Here the various densities are defined as
\begin{eqnarray}
 \rho_{\rm s}({\bf x}) &=& \sum_{\alpha}^{\text{occ}}
              \overline \psi_{\alpha}({\bf x})\psi_{\alpha}({\bf x})
          = \sum_a^{\text{occ}} {2j_a+1 \over 4\pi r^2}
          \Bigl( G_a^2(r) - F_a^2(r) \Bigr)
                           \ , \\[4pt]
 \rho_{\rm B}({\bf x}) &=& \sum_{\alpha}^{\text{occ}}
         \psi^\dagger_{\alpha}({\bf x})\psi_{\alpha}({\bf x})
          = \sum_a^{\text{occ}} {2j_a+1 \over 4\pi r^2}
          \Bigl( G_a^2(r) + F_a^2(r) \Bigr)
                           \ , \\[4pt] 
 \rho_{\rm B}^{\rm T}({\bf x}) &=&
        \sum_{\alpha}^{\text{occ}}
          \psi^\dagger_{\alpha}({\bf x})
           i\beta \bbox{\alpha\cdot\widehat r}
           \psi_{\alpha}({\bf x})
          = \sum_a^{\text{occ}} {2j_a+1 \over 4\pi r^2}
              2 G_a(r) F_a(r)
                           \ , \\[4pt]
 \rho_3({\bf x}) &=& \sum_{\alpha}^{\text{occ}}
         \psi^\dagger_{\alpha}({\bf x})\tau_3\psi_{\alpha}({\bf x})
          = \sum_a^{\text{occ}} {2j_a+1 \over 4\pi r^2} (2 t_a)
          \Bigl( G_a^2(r) + F_a^2(r) \Bigr)
                           \ , \\[4pt]
 \rho_3^{\rm T}({\bf x}) &=&
        \sum_{\alpha}^{\text{occ}}
          \psi^\dagger_{\alpha}({\bf x})
           i\tau_3\beta \bbox{\alpha\cdot\widehat r}
           \psi_{\alpha}({\bf x})  
          = \sum_a^{\text{occ}} {2j_a+1 \over 4\pi r^2} (2 t_a)
             2 G_a(r) F_a(r)
           \ ,
\end{eqnarray}
where the summation superscript ``occ'' means that the sum runs only over
occupied (valence) states in the Fermi sea.
Evidently, these densities depend only on the radial coordinate.
The charge density is given by
\begin{equation}
 \rho_{\rm chg}({\bf x}) \equiv\rho_{\rm d}({\bf x})+\rho_{\rm m}({\bf x})
                            \ ,
\end{equation}
where the ``direct'' nucleon charge density is
\begin{equation}
 \rho_{\rm d}({\bf x}) =
           \rho_{\rm p}({\bf x})+
           {1\over 2M^2}\bbox{\nabla\,\cdot\,}
           (\rho_{\rm a}^{\rm T}({\bf x})\widehat r)
           +{1\over 2M^2}[ \beta_{\rm s}
            \bbox{\nabla}^2\rho_{\rm B}+\beta_{\rm v}
                 \bbox{\nabla}^2\rho_3 ]
                            \ , \label{eq:direct}
\end{equation}
and the vector meson contribution is
\begin{equation}
\rho_{\rm m}({\bf x}) =
             {1\over g_\gamma g_\rho}\bbox{\nabla}^2R
           +{1\over 3g_\gamma \gomega}\bbox{\nabla}^2W \ .
\end{equation}
Here the ``point'' proton density and nucleon tensor density are determined
(as always) from the solutions to the Dirac equation in the mean fields:
\begin{eqnarray}
  \rho_{\rm p}({\bf x}) &=& {1\over 2} \sum_{\alpha}^{\text{occ}}
         \psi^\dagger_{\alpha}({\bf x})
            (1+\tau_3)\psi_{\alpha}({\bf x}) 
            = {1\over 2}(\rho_{\rm B} + \rho_3) \ , \label{eq:rhoppoint} \\
    \rho_{\rm a}^{\rm T}({\bf x}) &=&
             \sum_{\alpha}^{\text{occ}}
          \psi^\dagger_{\alpha}({\bf x})
           i\lambda \beta \bbox{\alpha\,\cdot\,}\widehat r
           \psi_{\alpha}({\bf x}) \ ,               
\end{eqnarray}
with the anomalous magnetic moment $\lambda$ from Eq.~(\ref{eq:lambda}).
%
%
%
As seen in the next section, the vector meson contribution and the final
term in Eq.~(\ref{eq:direct}) generate the nucleon form factor that reduces
the fluctuations of the point-nucleon charge densities.
Note that the mean-field equations can be used to eliminate the second
derivatives in the charge density in favor of first derivatives, which
are more accurately calculated numerically.

The energy functional is given by
\begin{equation}
  E = \sum_{\alpha}^{\text{occ}} E_{\alpha}
        -\int {\rm d}^3 x \, U_{\rm m} \ ,
\end{equation}
where
\begin{eqnarray}
 U_{\rm m} &\equiv& -{1\over 2}\Phi \rho_{\rm s}
              +{1\over 2}W(\rho_{\rm B}+\rho_{\rm B}^{\rm T})
              +{1\over 4}R(\rho_3+\rho_3^{\rm T})
              +{1\over 2}A\rho_{\rm d} \nonumber \\[4pt]
   & &   +{m_{\rm s}^2\over g_{\rm s}^2}{\Phi^3\over M}
         \Big({\kappa_3\over 12}+{\kappa_4\over 24}{\Phi\over M}
               +{\kappa_5\over 80}{\Phi^3\over M^3}\Big)
           -{\eta_\rho \over 4}{\Phi\over M} 
                { m_\rho^2\over g_\rho^2} R^2
             \nonumber \\[4pt]
   & & -{\Phi \over 4M}
         \Big(\eta_1+\eta_2{\Phi\over M}
               +{\eta_3\over 2}{\Phi^2\over M^2}\Big)
            { \momega^2 \over \gomega^2} W^2
                \nonumber \\[4pt]
   & & -{1\over 4! \gomega^2}
         \Big(\zeta_0 +{3\zeta_1\over 2}{\Phi\over M}\Big) W^4
       +{\alpha_1\over 4g_{\rm s}^2}{\Phi\over M}
             (\bbox{\nabla}\Phi)^2
       - {\alpha_2\over 4\gomega^2}{\Phi\over M}
            (\bbox{\nabla}W)^2 \ ,
\end{eqnarray}
and we have applied the mean-field equations to arrive at this form.

The center-of-mass (c.m.) 
correction to the nuclear binding energy can be estimated 
nonrelativistically using
\begin{equation}
  E_{\rm CM} = {\langle {\hat{P}}_{\rm CM}^2\rangle
                   \over 2 M B} \ , \label{eq:Ecm}
\end{equation}
where $B=Z+N$ is the baryon number of the nucleus, and \cite{REINHARD89}
\begin{equation}
  \langle {\hat{P}}_{\rm CM}^2\rangle \equiv
       -\sum_{\alpha}^{\text{occ}} 
        \langle \alpha |\bbox{\nabla}^2|\alpha\rangle
        +\sum_{\alpha ,\beta}^{\text{occ}}
        |\langle \alpha |\bbox{\nabla}|\beta\rangle|^2 \ ,
\end{equation}
with 
\begin{equation}
\langle \alpha |\hat{O}|\beta\rangle \equiv
        \int {\rm d}^3 x\, 
        \psi^\dagger_{\alpha}({\bf x})\hat{O}\psi_{\alpha}({\bf x})\ .
\end{equation}
For simplicity, we use an empirical
estimate given by Reinhard\cite{REINHARD89}:
\begin{equation}
  E_{\rm CM} = 17.2 /B^{1/5} \ ( {\rm MeV} ) \ ,
\end{equation}
which determines ${\langle {\hat{P}}_{\rm CM}^2\rangle}$ as well from
Eq.~(\ref{eq:Ecm}).
The binding energy $\epsilon$ is then
\begin{equation}
  \epsilon = B M - (E - E_{\rm CM}) \ .
\end{equation}
Note that we cannot replace $B M$ by 
$Z m_{\rm p} + N m_{\rm n}$, the sum of
the masses of the protons and neutrons, unless we use different
masses for the proton and neutron in the Dirac equation.

There are also c.m.~corrections to the nuclear charge density and charge 
radius.
We estimate these in analogy to the well-known nonrelativistic results for
harmonic-oscillator wavefunctions, which are exact.
The charge density in momentum space, {\em i.e.}, the charge form factor, 
becomes
\begin{equation}
 F_{\rm chg} ({\bf q}) = {1\over Z}{\rho_{\rm chg}}({\bf q})
              \Bigl(1 + {{\bf q}^2\over 8
              \langle {\hat{P}}_{\rm CM}^2\rangle } + \cdots \Bigr) \ , 
\end{equation}
where ${\rho_{\rm chg}}({\bf q})$ is the Fourier transform of the charge 
density $\rho_{\rm chg}({\bf x})$.
We expect this correction to be accurate for momenta
$|{\bf q}| \lesssim 2\, {\rm fm}^{-1}$.
With the c.m.~correction, the predicted charge density 
in coordinate space is, to a good approximation,
\begin{equation}
\{\rho_{\rm chg}({\bf x})\}_{\rm obs}\approx \rho_{\rm chg}({\bf x})
         -{1\over 8 \langle {\hat{P}}_{\rm CM}^2\rangle}
              \bbox{\nabla}^2\rho_{\rm chg}({\bf x})      \ .
\end{equation}
The mean-square charge radius is given by
\begin{equation}
\langle r^2 \rangle_{\rm chg} = \langle r^2 \rangle 
             -{3\over 4 \langle {\hat{P}}_{\rm CM}^2\rangle}    \ ,
\end{equation}
where
\begin{equation}
 \langle r^2 \rangle={1\over Z}\int {\rm d}^3 x\, 
           {\bf x}^2 \rho_{\rm chg}({\bf x})  \ .
\end{equation}

The energy density for uniform, symmetric nuclear matter in the one-loop
approximation can be obtained from the preceding results
by observing that the single-particle energy eigenvalue becomes
\begin{equation}
 E({\bf k})=W +\sqrt{{\bf k}^2+{M^*}^2} \ ,
\end{equation}
where $M^*=M-\Phi$, and $\Phi$ and $W$ are now constant
mean fields [see Eqs.~(\ref{eq:Phi}) and (\ref{eq:W})]. 
The energy density ${\cal E}$ becomes
\begin{eqnarray}
     {\cal E}[M^\ast,\rho_{{\scriptscriptstyle\rm B}}]
        &=& 
       {m_{\rm s}^2\over g_{\rm s}^2}\Phi^2
         \Big({1\over 2}+
               {\kappa_3\over 3!}{\Phi\over M}
              +{\kappa_4\over 4!}{\Phi^2\over M^2}
               +{\kappa_5\over 5!}{\Phi^3\over M^3}\Big)
                \nonumber  \\[4pt]
     & & \null -{1\over 2\gomega^2}
         \Big(1+\eta_1{\Phi\over M}+{\eta_2\over 2}{\Phi^2\over M^2}
               +{\eta_3\over 3!}{\Phi^3\over M^3}\Big)
                   \momega^2 W^2
          \nonumber  \\[4pt]
     & & \null
            - {1\over 4! \gomega^2}
              \Big(\zeta_0 +\zeta_1 {\Phi\over M}\Big )W^4
           +W\rho_{\scriptscriptstyle\rm B}
     + {4\over (2\pi)^3}\! \int^{k_{\scriptscriptstyle\rm F}}
       {\kern-.1em}{\rm d}^3{\kern-.1em}{k} \,
           \sqrt{{\bf k}^2+{M^*}^2} \ , \label{eq:nuclmatt}
\end{eqnarray}
where $k_{\scriptscriptstyle\rm F}$ is the Fermi momentum
and
$\rho_{\scriptscriptstyle\rm B}=
2 k_{\scriptscriptstyle\rm F}^3/3\pi^2$ is the baryon density.
An extensive analysis of energy densities of this form is given in
Ref.~\cite{bodmer}. 
One can also compute the bulk symmetry-energy coefficient \cite{SEROT86}
\begin{equation}
  a_4 = {g_\rho^2 \over 12\pi^2 {m_\rho^*}^2}
         k_{\rm \scriptscriptstyle  F}^3
          + {1 \over 6} {k_{\rm \scriptscriptstyle  F}^2
          \over \sqrt{k_{\rm \scriptscriptstyle  F}^2+{M^*}^2}} \ ,
\end{equation}
where the effective rho mass $m_\rho^*$ is defined here by 
\begin{equation} 
{m_\rho^*}^2 \equiv m_\rho^2 (1 + \eta_\rho \Phi/M)  \ .
\end{equation}

\section{Results}

To test the physical ideas discussed in the preceding sections,
we fit our one-loop results to the properties of finite nuclei.
We choose a set of spherical nuclei and solve the 
Dirac and mean-field equations for each nucleus
by iteration, until self-consistency is reached \cite{HOROWITZ}. 
We then calculate a set of
observables \{$X_{\rm th}^{(i)}$\} for each nucleus
and tune the parameters to minimize the generalized $\chi^2$
\cite{LosAlamos} defined by
\begin{equation}
\chi^2 = \sum_{i} \sum_{X}
               \bigg[{ X_{\rm exp}^{(i)}-X_{\rm th}^{(i)}
                      \over W_{X}^{(i)} X_{\rm exp}^{(i)}} 
                \bigg]^2 \ ,  \label{eq:chisq}
\end{equation}
where $i$ runs over the set of nuclei, $X$ runs over the set of
observables, the subscript ``exp'' indicates the experimental value
of the observable, and $W_{X}^{(i)}$ are the relative weights.
The nuclei we choose are $^{16}$O, $^{40}$Ca, $^{48}$Ca, 
$^{88}$Sr, and $^{208}$Pb. 

Similar fitting procedures have been standard for determining parameter sets
for relativistic mean-field 
models \cite{REINHARD89,LosAlamos,RUFA88,SHARMA93}.
In the previous studies, however, the principal goal was to determine
minimal parameter sets that provided a good description of bulk nuclear
properties, so as to maximize predictability.
The set of observables used to make the fit was also kept at a minimum.
In these fits, the parameters included only
the scalar mass $m_{\rm s}$, the nucleon-meson couplings
$g_{\rm s}$, $g_{\rm v}$, and $g_{\rho}$, and two scalar self-interaction
parameters, $\kappa_3$ and $\kappa_4$.
(The exception is the point coupling model of Ref.~\cite{LosAlamos},
which has a corresponding set of nucleon-only point couplings.)
The nucleon and vector meson masses were generally fixed 
at experimental values.
The resulting parameter sets provide good fits to nuclear
binding energies, charge distributions, and single-particle structure,
for a wide variety of nuclei.

Our effective lagrangian, truncated at $\nu=4$, contains
many additional parameters.  
Thus we  expect to find good fits to nuclear properties.
However, our goals are quite different from the previous studies.
We are not concerned with maximizing predictive power; instead,
we want to use 
experimental data to test whether the parameters of the effective lagrangian
are, in fact, natural.  
In so doing 
we want to gain insight into an appropriate level of truncation.
Therefore we 
constrain the parameters by using every
observable we believe should be well reproduced.
 
The $\chi^2$ function in Eq.~(\ref{eq:chisq}) serves as a figure of merit
for our fit.  We do not expect our model to reproduce the experimental
observables at the level of the actual experimental uncertainties in each case;
this means that a direct statistical interpretation of $\chi^2$
will not be meaningful.
Rather we choose the weights to be the relative accuracy we expect
in a good fit. 
A ``perfect fit'' corresponds to a contribution of unity to $\chi^2$ from
each observable. 
In choosing weights we follow closely the prescription in 
Ref.~\cite{LosAlamos}, so that we will have a basis of comparision. 
Different choices of weights will emphasize different aspects of
the physics and lead to different ``optimal'' parametrizations.
However, we have tried a range of reasonable weights and the qualitative
conclusions given below are independent.

The choice of observables starts with
the basic physics of the semi-empirical mass formula, which 
should be reproduced by any reasonable model of nuclei.
Here this is achieved directly
by fitting to the binding energies of individual nuclei.
We adopt a relative weight of 0.15\% for each binding energy.
This implies that binding energies per nucleon should be reproduced to two
decimal places, with an uncertainty of roughly unity in the second
place.
But it is also important that the {\it systematics\/} of the energies
are followed; otherwise, the minimization procedure
may find reasonable fits
to energies but with poor systematics (particularly when fewer parameters
are used, as with $\nu=2$ or $\nu=3$ truncations).  
Therefore we also use the deviation of experimental
and calculated binding energies 
$\delta\epsilon_i \equiv (\epsilon_i)_{\rm exp} - (\epsilon_i)_{\rm th}$
to determine $\delta a_1$, $\delta a_2$,
and $\delta a_4$ according to
\begin{equation}
    \delta \epsilon_i = \delta a_1\, A_i - \delta a_2\, A_i^{2/3}
         - \delta a_4\, (N_i-Z_i)^2/A_i  \ .  \label{eq:SEMF}
\end{equation}
Here $N_i$ and $Z_i$ are the number of neutrons and protons in the
$i^{\rm th}$ nucleus and $A_i = N_i + Z_i$.
Thus an exact fit to the energies would also have $\delta a_1 =
\delta a_2 = \delta a_4 = 0$.
The deviations $\delta a_2$ and $\delta a_4$ are included as separate
terms in Eq.~(\ref{eq:chisq}) in the form
$[\delta a_i/W_{\delta a_i}]^2$, 
with $W_{\delta a_i}=0.08$.  
(This weight was selected based on
the results for low-order truncations; it has no intrinsic significance.)

We also expect 
the effective theory to accurately reproduce low-momentum observables,
which leads
us to focus on the charge form factor.
We choose the rms charge radii of the nuclei and the so-called
diffraction-minimum-sharp (d.m.s.) radii as observables
(another choice would have been the surface thickness \cite{RUFA88}).
The d.m.s.\ radius of a nucleus is defined to be\cite{FRIEDRICH82}
\begin{equation}
 R_{\rm dms}\equiv 4.493/Q_0^{(1)}  \ ,
\end{equation}
where $Q_0^{(1)}$ is the 
three-momentum transfer at the first zero of the nuclear 
charge form factor $F(Q) \equiv F_{\rm chg}({\bf q})$ with 
$Q=|\bf q|$.
The d.m.s.\ radius is expected to be 
a clean low-momentum quantity for a nucleus and should be well described.
The weights chosen for the radii correspond to an accuracy of
roughly two decimal places.

Finally, we expect to reproduce various features of the single-particle
structure in  nuclei.  We focus in particular
on spin-orbit splittings near the Fermi surface, which
are reasonably well known for both proton and neutron levels.
These splittings are closely related to the size of the mean fields
at the one-loop level.
In a previous study, a good reproduction of these splittings was found
even when they were not included as observables \cite{RUFA88}.
However, our expanded parameter space means that this correlation may not
follow without explicitly including them.
In addition, we use
the proton $1h_{9/2}$ energy level in $^{208}$Pb 
to fix the overall scale of the  energy levels;
the positioning of this level helps to determine parameters that
influence the symmetry energy.
Finally, the proton splitting 
$E_{\rm p}(2d_{3/2})-E_{\rm p}(1h_{11/2})$ 
is chosen as a characteristic splitting between states with large and
small $l$, which constrains the shape of the effective single-particle
potential \cite{FETTER71}.
The additional weight on observables for lead also leads to
a more sensitive determination of the isovector parameters.

To summarize,
the observables (a total of 29) 
and their relative weights are taken as follows:
\begin{itemize}
\item The binding energies per nucleon $\epsilon /B$, with a relative
       weight of 0.15\%
\item The rms charge radii
        $\langle r^2 \rangle^{1\over 2}_{\rm chg}$, with
         a relative  weight of  $0.2\%$
\item The d.m.s. radii $R_{\rm dms}$,
         with a relative weight of  $0.15\%$
\item The spin-orbit splittings $\Delta E_{\rm SO}$ of the
      least-bound protons and neutrons, with
         a relative  weight of $5\%$ for $^{16}$O, $15\%$ for
     $^{208}$Pb,
         $25\%$ for $^{40}$Ca and $^{48}$Ca, and
         $50\%$ for $^{88}$Sr
\item The proton energy  $E_{\rm p}(1h_{9/2})$ and the
          proton level splitting
          $E_{\rm p}(2d_{3/2}) - E_{\rm p}(1h_{11/2})$
          in $^{208}$Pb, with relative
          weights of  $5\%$ and $25\%$, respectively
\item The surface-energy and symmetry-energy
    deviation coefficients $\delta a_2$ and $\delta a_4$
    [see Eq.~(\ref{eq:SEMF})], each with
        a  weight of $0.08$.
\end{itemize}
For further discussion of the relative weights for $\epsilon /B$,
$\langle r^2\rangle^{1/2}_{\rm ch}$, and $\Delta E_{\rm SO}$, see 
Ref.\cite{LosAlamos}.
Additional observables that might be added in future investigations 
include 
neutron radii, quadrupole deformations, monopole resonance energies,
and an enlarged set of test nuclei.

We take the nucleon, $\omega$, and $\rho$ masses to be given
by their experimental values:
$M=939\,$MeV, $\momega = 782\,$MeV, and $m_{\rho}=770\,$MeV.%
\footnote{Including $\momega$ as a free parameter starting from the parameter
sets in Table~\ref{tab:one}
leads to only a tiny improvement in $\chi^2$ (less than unity) 
and a small change
in $\momega$ (about 10~MeV).  
Therefore we fix the $\omega$ and $\rho$ masses, 
since the latter also has little effect.}
The anomalous magnetic moments of the nucleon are fixed at
$\lambda_{\rm p}=1.793$ and $\lambda_{\rm n}=-1.913$.
We also fix $g_\gamma=5.01$ as given in Eq.~(\ref{eq:ggamma}).
The free-space charge radii of the nucleon, as given in 
Eqs.~(\ref{eq:nucsize1}) and (\ref{eq:nucsize2}), are used
to fix $\beta_{\rm s}$, $\beta_{\rm v}$, and $f_\rho$ by solving
Eqs.~(\ref{eq:size1}) to (\ref{eq:size2}). 
The remaining thirteen parameters $g_{\rm s}$, $\gomega$,
$g_{\rho}$, $\eta_1$, $\eta_2$, $\eta_\rho$, $\kappa_3$, $\kappa_4$,
$\zeta_0$, $m_{\rm s}$, $\fomega$, $\alpha_1$, and $\alpha_2$
for the $\nu=4$ parametrization
are then obtained by the optimization procedures described above.

This optimization program faces serious difficulties.
With such a large parameter set and because the observables themselves
are highly correlated, there is a definite problem of
underdetermination.
The minimization routines used here are reliable for determining
local minima but are not guaranteed to find a global minimum,
so we cannot be sure to have sampled all relevant values of
the parameters.
There are further complications in obtaining a more global fit because
the required stability of the Hartree
iterations precludes large steps through parameter space.
Thus the minimization routines 
must be efficient in navigating narrow valleys in
parameter space.
To make our results as robust as possible, we used two different
codes and made many runs with different starting parameters.

\def\mc#1{\multicolumn{1}{c}{$\quad #1$}}
\def\zz{\phantom{0}}

\begin{table}[t]
\caption{Parameter sets
from fits to finite nuclei.
}
\vspace{.1in}
\begin{tabular}[t]{crrrrrr}
                 & \mc{W1} & \mc{C1} & \mc{Q1} & \mc{Q2} & \mc{G1} & \mc{G2} \\
   \hline
$m_{\rm s}/M$    & 0.60305 & 0.53874 & 0.53735 & 0.54268 & 0.53963 & 0.55410   \\     
$g_{\rm s}/4\pi$ & 0.93797 & 0.77756 & 0.81024 & 0.78661 & 0.78532 & 0.83522   \\     
$\gomega/4\pi$   & 1.13652 & 0.98486 & 1.02125 & 0.97202 & 0.96512 & 1.01560   \\     
$g_\rho/4\pi$    & 0.77787 & 0.65053 & 0.70261 & 0.68096 & 0.69844 & 0.75467   \\     
$\eta_1$         &         & 0.29577 &         &         & 0.07060    & 0.64992   \\     
$\eta_2$         &         &         &         &         & $-$0.96161 & 0.10975   \\     
$\kappa_3$       &         & 1.6698\zz  & 1.6582\zz  & 1.7424\zz  & 2.2067\zz     & 3.2467\zz    \\     
$\kappa_4$       &         &     & $-$6.6045\zz & $-$8.4836\zz & $-$10.090\zz\zz  & 0.63152   \\     
$\zeta_0$        &         &         &         & $-$1.7750\zz & 3.5249\zz     & 2.6416\zz    \\     
$\eta_\rho$      &         &         &         &        & $-$0.2722\zz  & 0.3901\zz    \\     
$\alpha_1$       &         &         &         &        & 1.8549\zz     & 1.7234\zz    \\     
$\alpha_2$       &         &         &         &        & 1.7880\zz     & $-$1.5798\zz \\     
$\fomega/4$      &         &         &         &        & 0.1079\zz     & 0.1734\zz    \\
 \hline     
$f_{\rho}/4$     & 0.9332\zz & 1.1159\zz & 1.0332\zz & 1.0660\zz & 1.0393\zz     & 0.9619\zz    \\     
$\beta_{\rm s}$  & $-$0.38482 & $-$0.01915 & $-$0.10689 & 0.01181 & 0.02844 & $-$0.09328 \\     
$\beta_{\rm v}$  & $-$0.54618 & $-$0.07120 & $-$0.26545 & $-$0.18470 & $-$0.24992    & $-$0.45964     
\end{tabular}
\label{tab:one}
\end{table}

\begin{figure}
 \setlength{\epsfxsize}{4.0in}
  \centerline{\epsffile{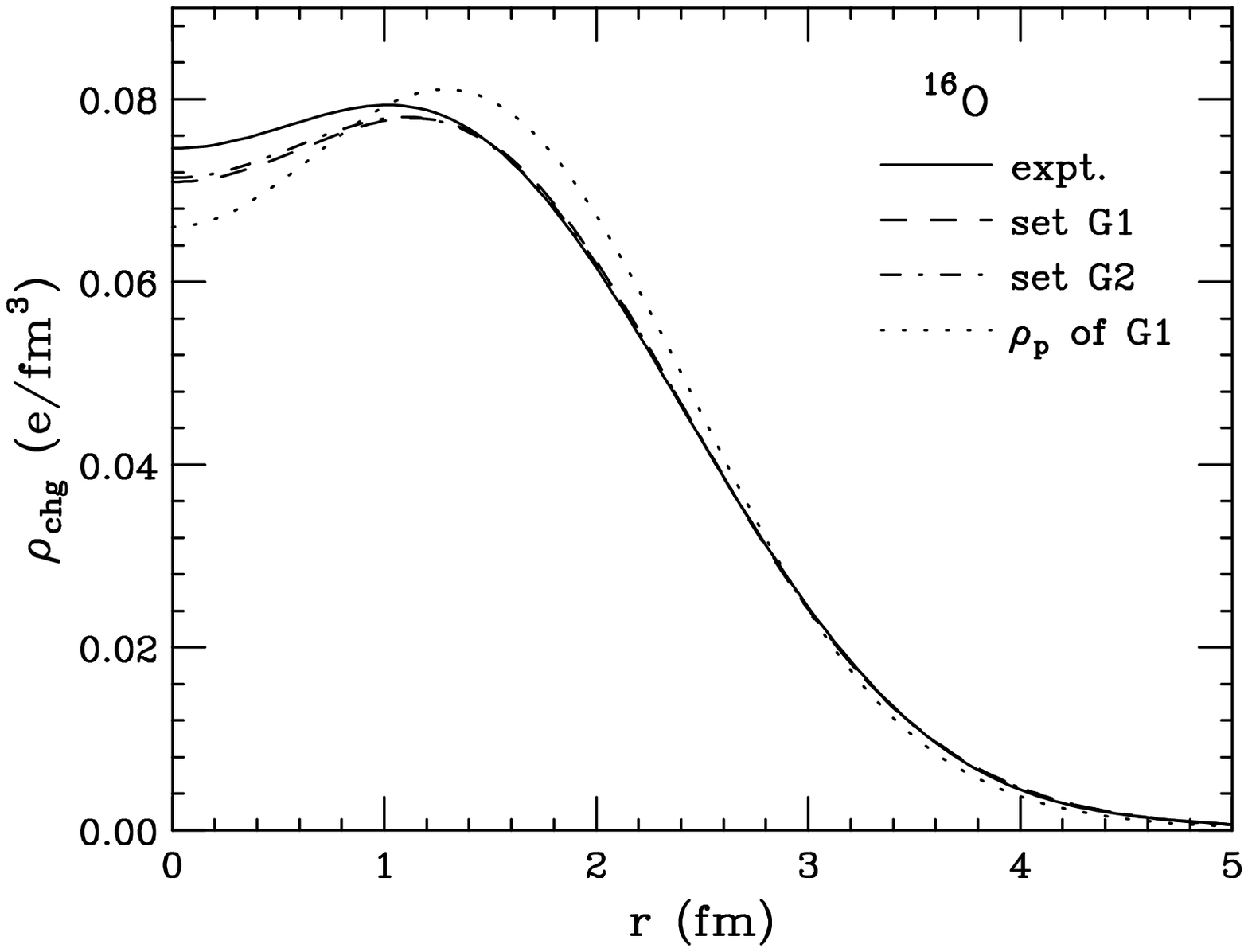}}
\vspace{.2in} \caption{Charge density of $^{16}$O.
The solid line is taken from experiment \protect\cite{DEVRIES87}.
Charge densities are shown for the G1 and G2 parameter sets
from Table~I. Also shown is the point-proton 
density [$\rho_{\rm p}$ of Eq.~(\ref{eq:rhoppoint})] for set G1.}
 \label{fig:one}
\end{figure}

\begin{figure}
 \setlength{\epsfxsize}{4.0in}
  \centerline{\epsffile{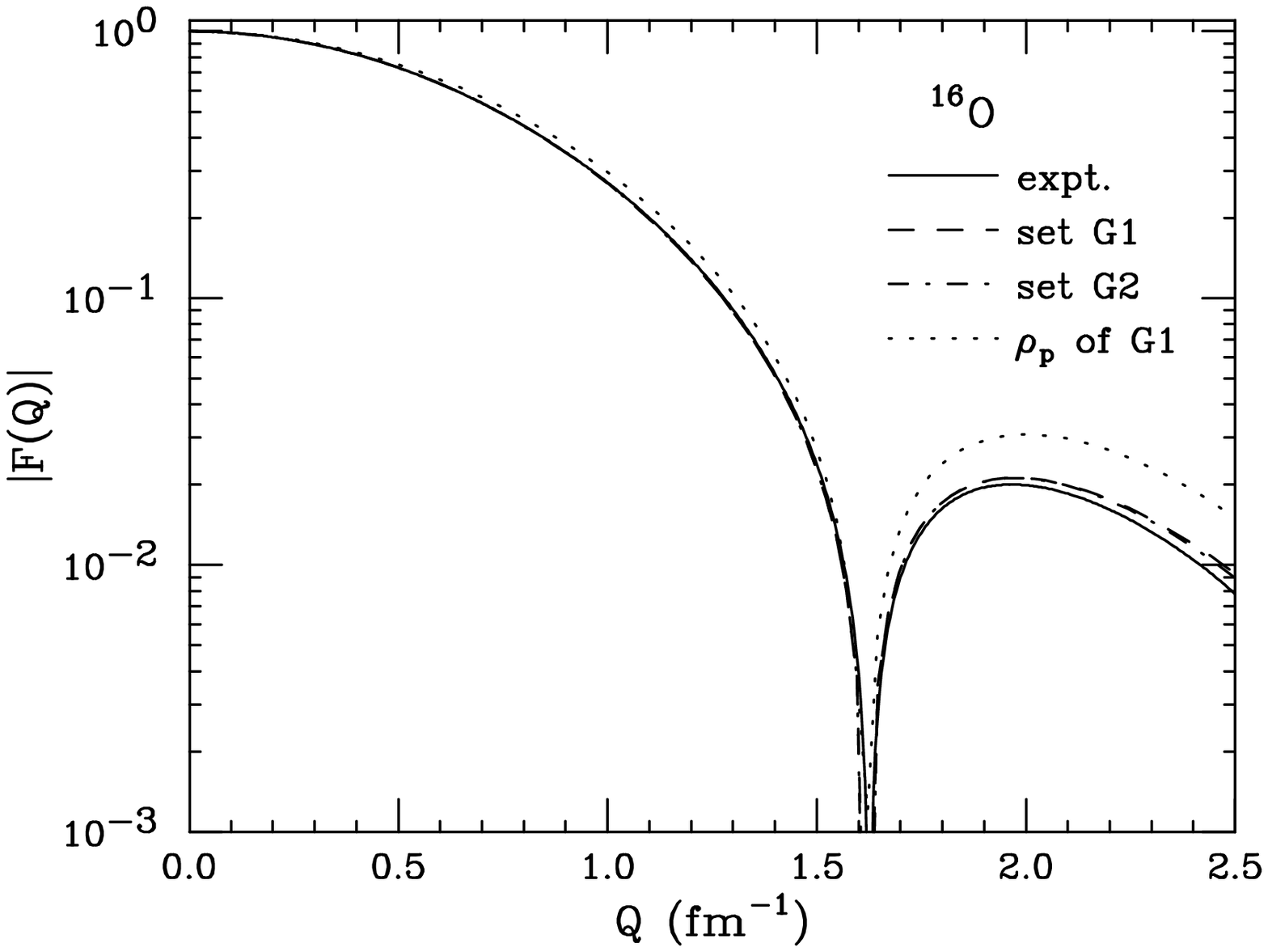}}
\vspace{.2in} \caption{Charge form factor of $^{16}$O. 
The solid line is taken from experiment \protect\cite{DEVRIES87}.
Form factors are shown for the G1 and G2 parameter sets
from Table~I.  }
\label{fig:two}
\end{figure}

\begin{figure}
 \setlength{\epsfxsize}{4.0in}
  \centerline{\epsffile{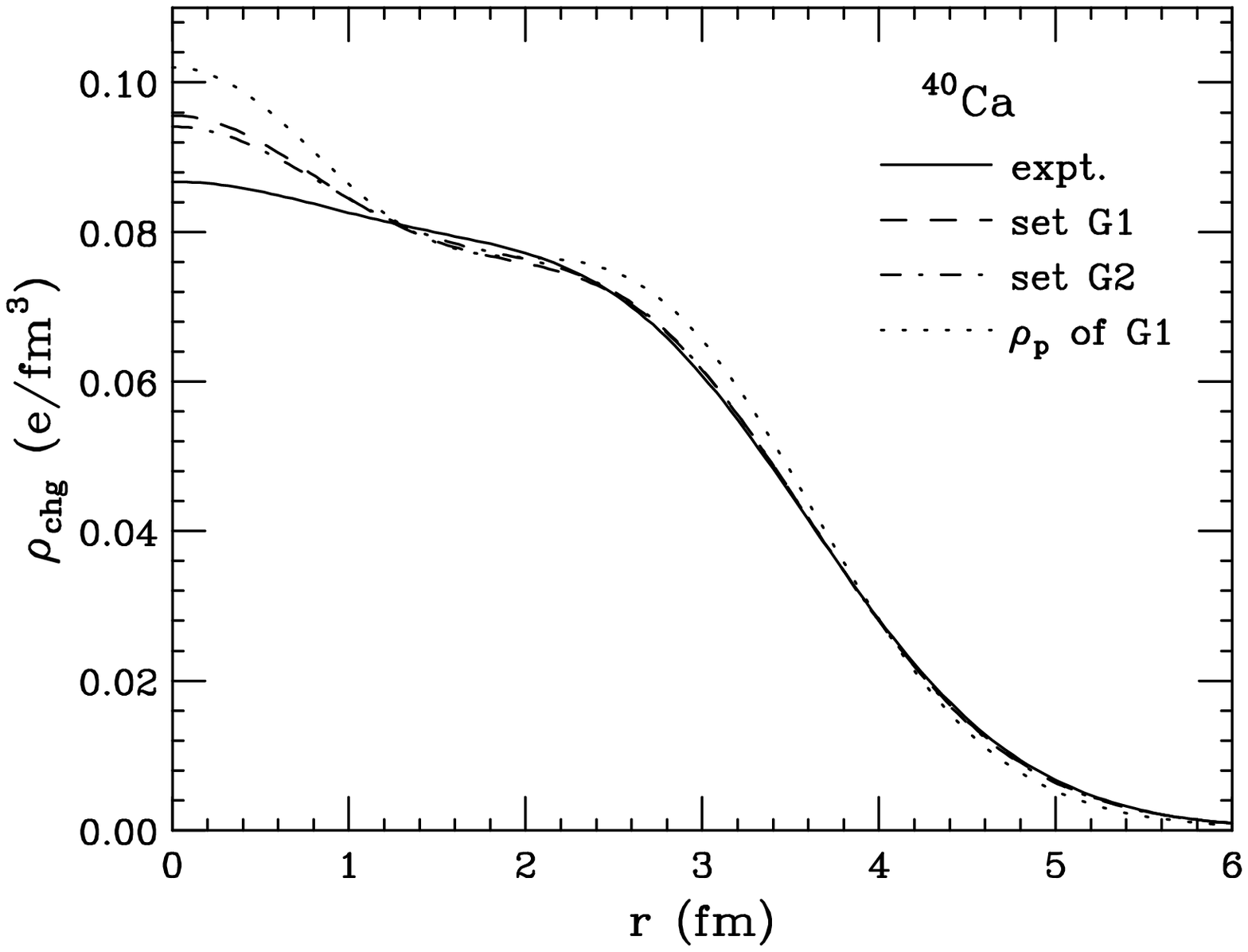}}
\vspace{.2in} \caption{Charge density of $^{40}$Ca.
The solid line is taken from experiment \protect\cite{DEVRIES87}.
Charge densities are shown for the G1 and G2 parameter sets
from Table~I. Also shown is the   
point-proton density for set G1.}
 \label{fig:three}
\end{figure}

\begin{figure}
 \setlength{\epsfxsize}{4.0in}
     \centerline{\epsffile{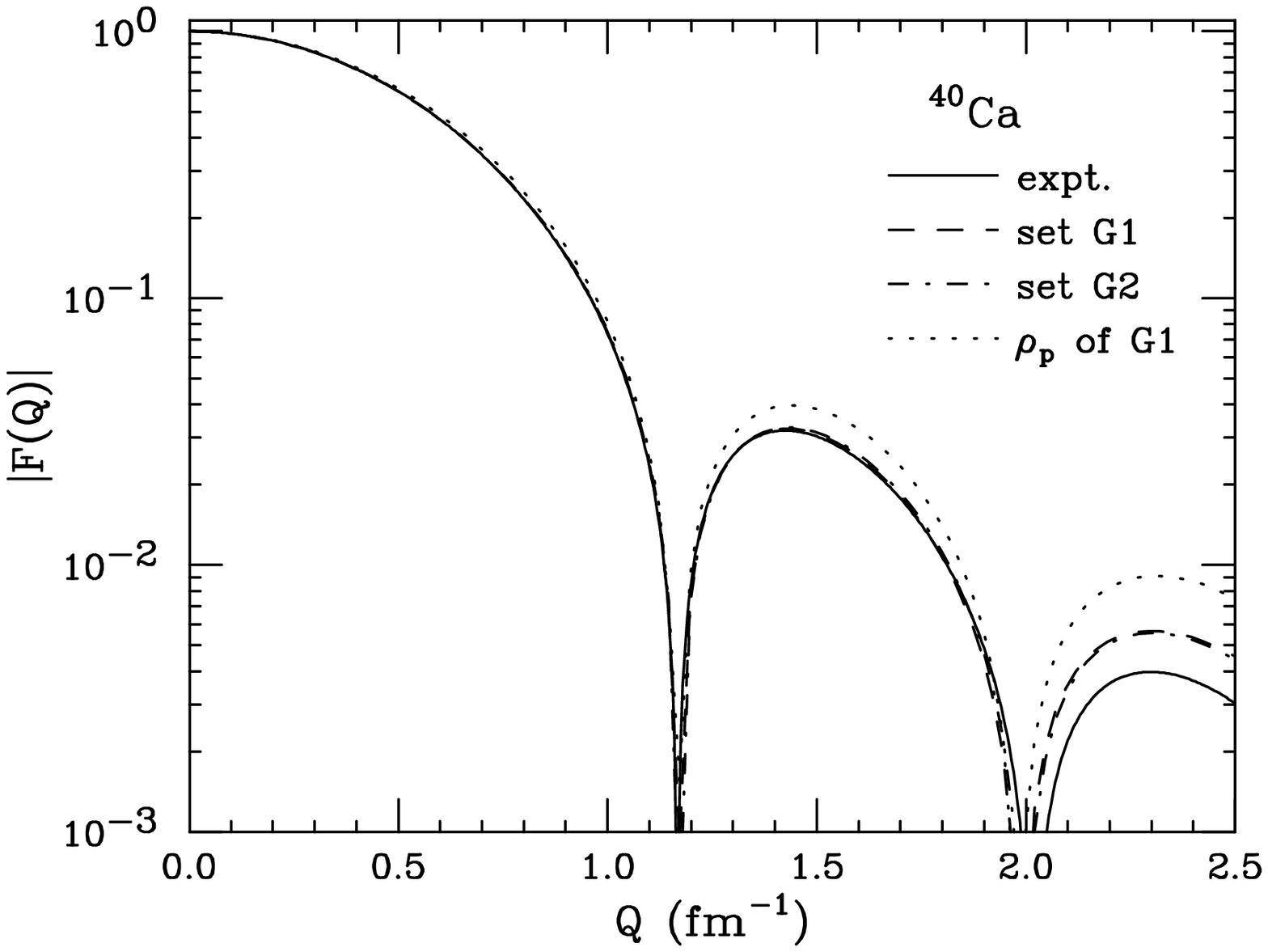}}
\vspace{.2in} \caption{Charge form factor of $^{40}$Ca.
The solid line is taken from experiment \protect\cite{DEVRIES87}.
Form factors are shown for the G1 and G2 parameter sets
from Table~I.  }
 \label{fig:four}
\end{figure}

\begin{figure}
 \setlength{\epsfxsize}{4.0in}
  \centerline{\epsffile{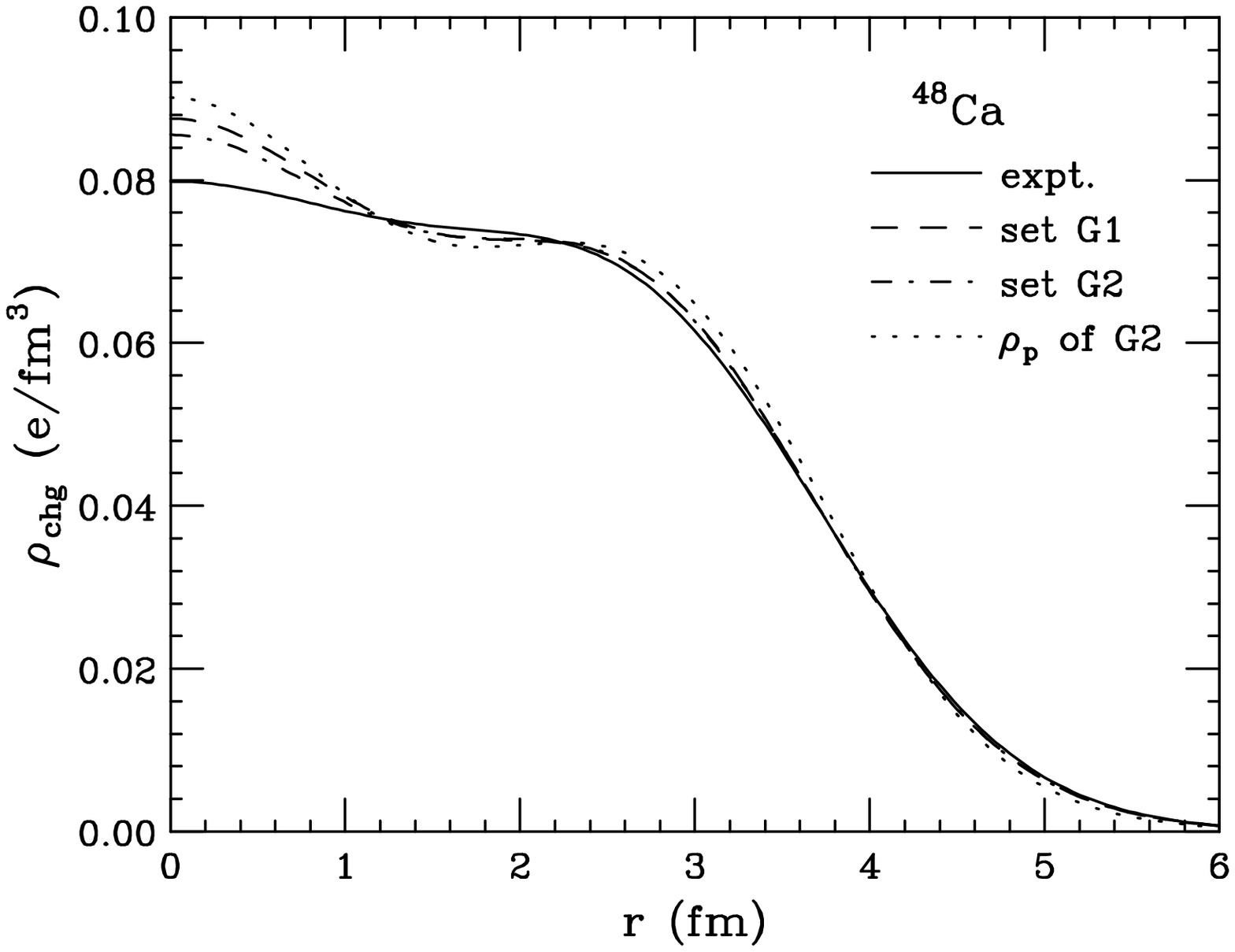}}
\vspace{.2in} \caption{Charge density of $^{48}$Ca.
The solid line is taken from experiment \protect\cite{DEVRIES87}.
Charge densities are shown for the G1 and G2 parameter sets
from Table~I. Also shown is the  
point-proton density for set G2.}
 \label{fig:five}
\end{figure}

\begin{figure}
 \setlength{\epsfxsize}{4.0in}
     \centerline{\epsffile{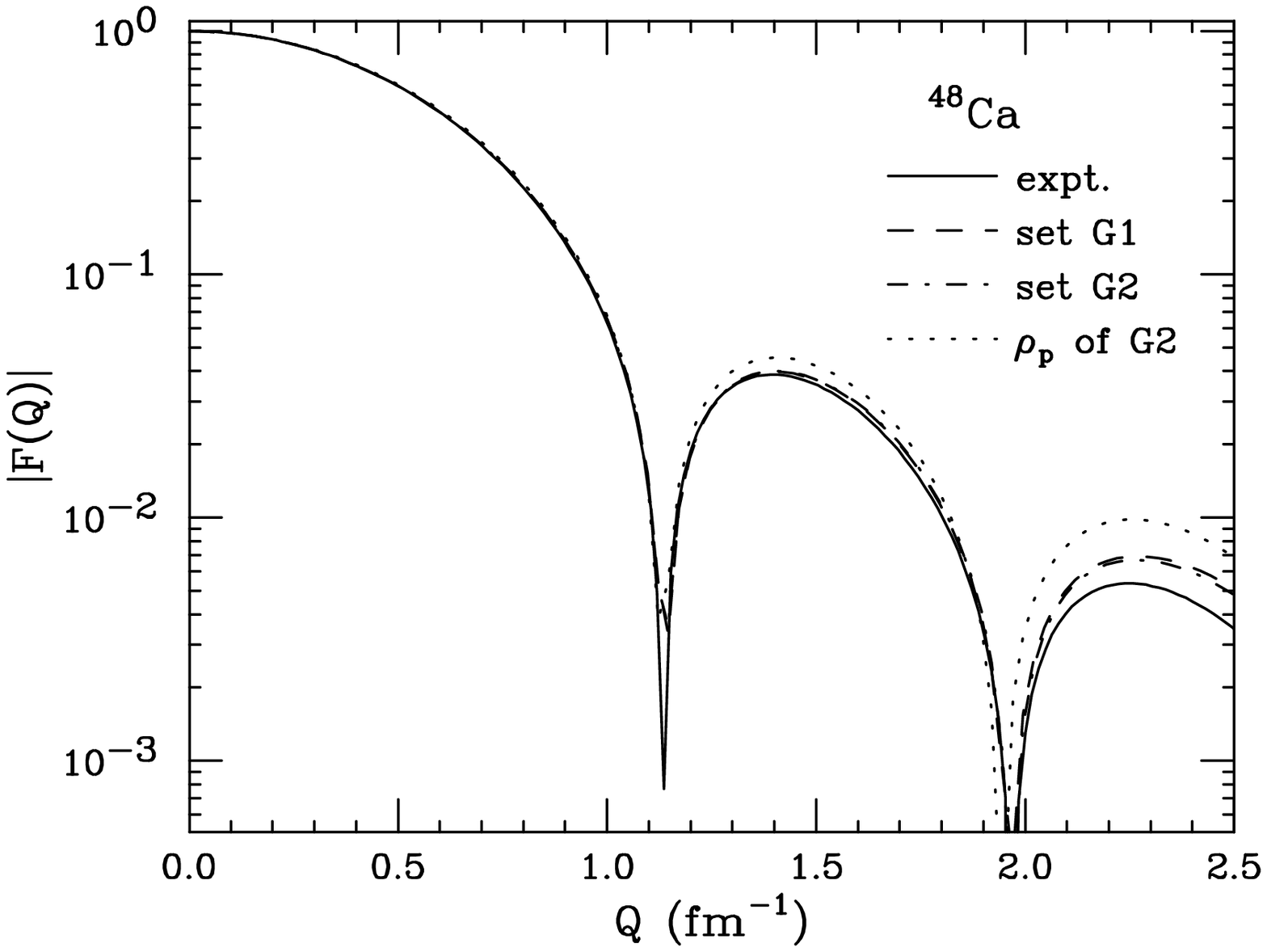}}
\vspace{.2in} \caption{Charge form factor of $^{48}$Ca.
The solid line is taken from experiment \protect\cite{DEVRIES87}.
Form factors are shown for the G1 and G2 parameter sets
from Table~I.  }
 \label{fig:six}
\end{figure}

\begin{figure}
 \setlength{\epsfxsize}{4.0in}
 \centerline{\epsffile{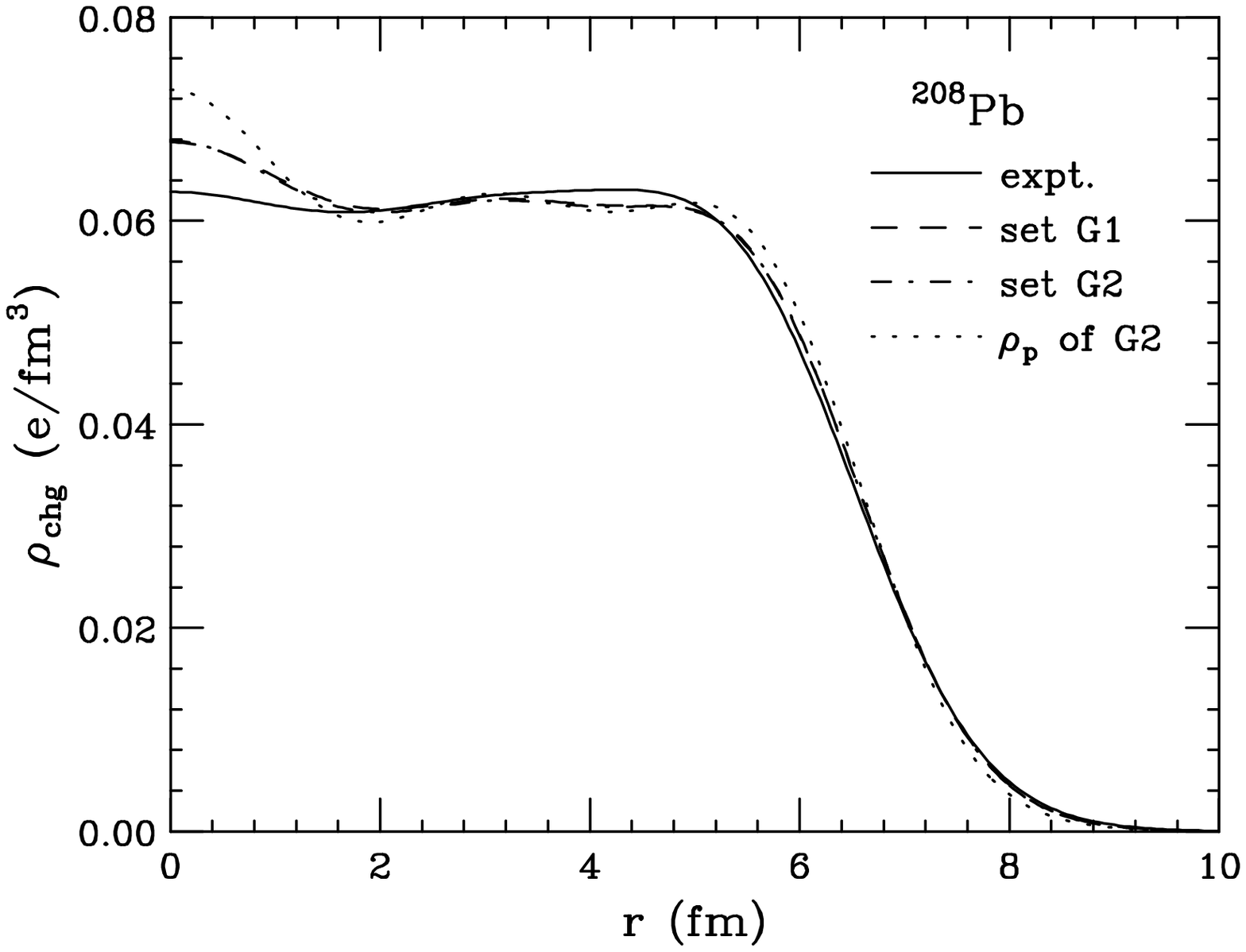}}
\vspace{.2in} \caption{Charge density of $^{208}$Pb.
The solid line is taken from experiment \protect\cite{DEVRIES87}.
Charge densities are shown for the G1 and G2 parameter sets
from Table~I. Also shown is the point-proton 
density for set G2.}
 \label{fig:nine}
\end{figure}

\begin{figure}
 \setlength{\epsfxsize}{4.0in}
 \centerline{\epsffile{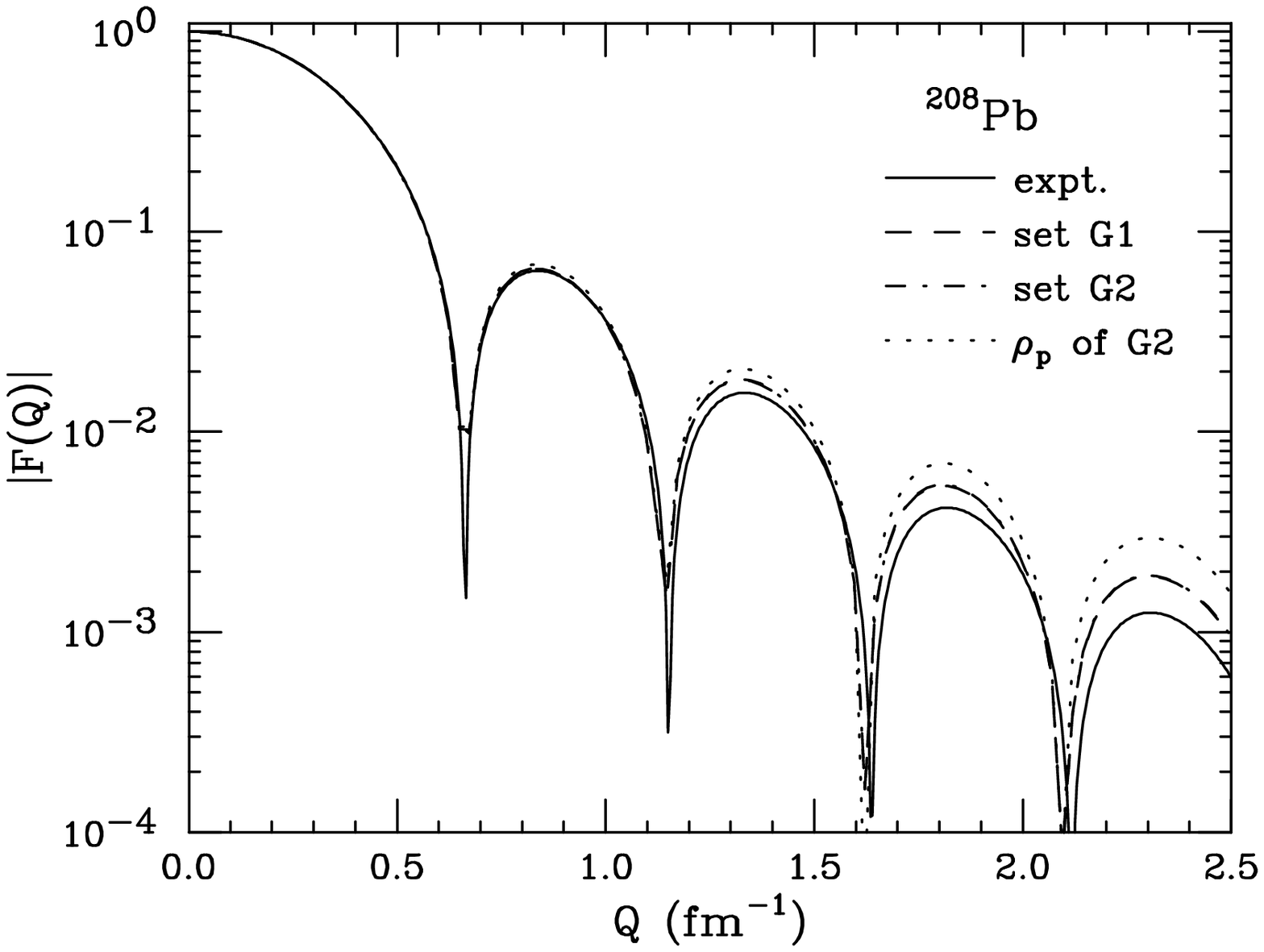}}
\vspace{.2in} \caption{Charge form factor of $^{208}$Pb.
The solid line is taken from experiment \protect\cite{DEVRIES87}.
Form factors are shown for the G1 and G2 parameter sets
from Table~I.  }
 \label{fig:ten}
\end{figure}

\begin{figure}[t]
 \setlength{\epsfxsize}{4.0in}
 \centerline{\epsffile{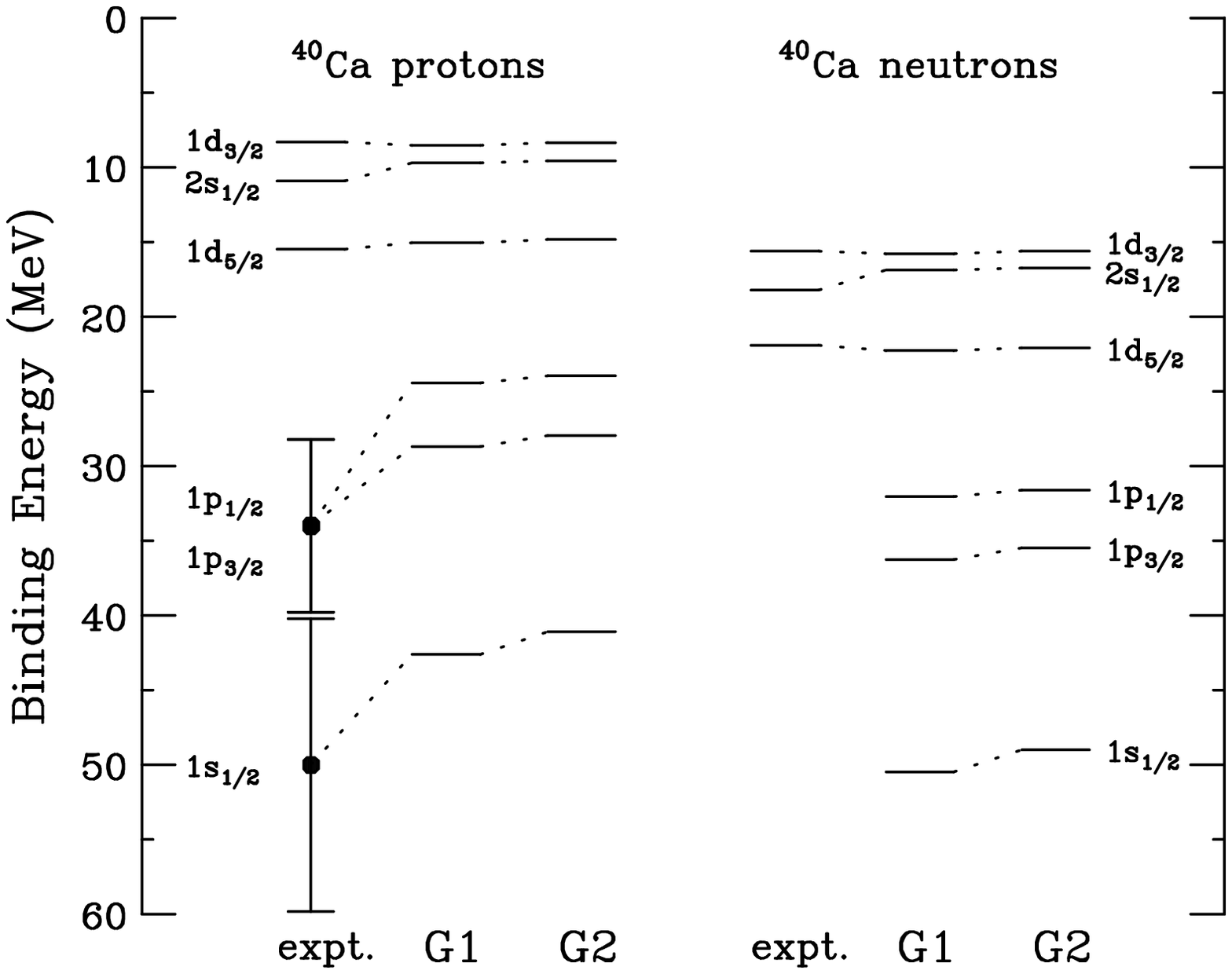}}
\vspace{.2in} 
\caption{
Predicted single-particle spectra for $^{40}$Ca using
the G1 and G2 parameter sets from Table~I.
The leftmost values are from experiment \protect\cite{CAMPI72}.}
\label{fig:calev}
\end{figure}

\begin{figure}[t]
 \setlength{\epsfxsize}{4.0in}
 \centerline{\epsffile{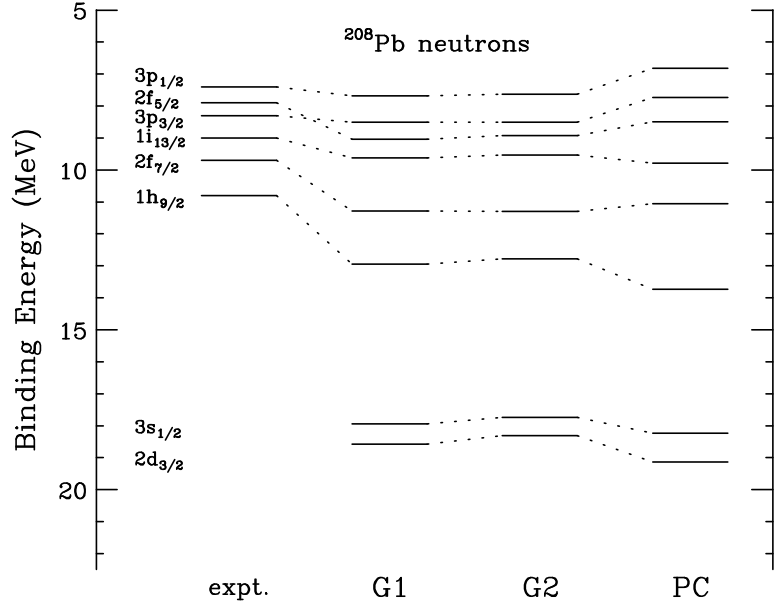}}
\vspace{.2in} 
\caption{
Predicted neutron single-particle spectra for $^{208}$Pb 
near the Fermi surface
using
the G1 and G2 parameter sets from Table~I.
The leftmost values are from experiment \protect\cite{CAMPI72}, 
and PC stands for 
the point-coupling
model of Ref.~\protect\cite{LosAlamos}.}
\label{fig:elevena}
\end{figure}

\begin{figure}[!b]
 \setlength{\epsfxsize}{4.0in}
 \centerline{\epsffile{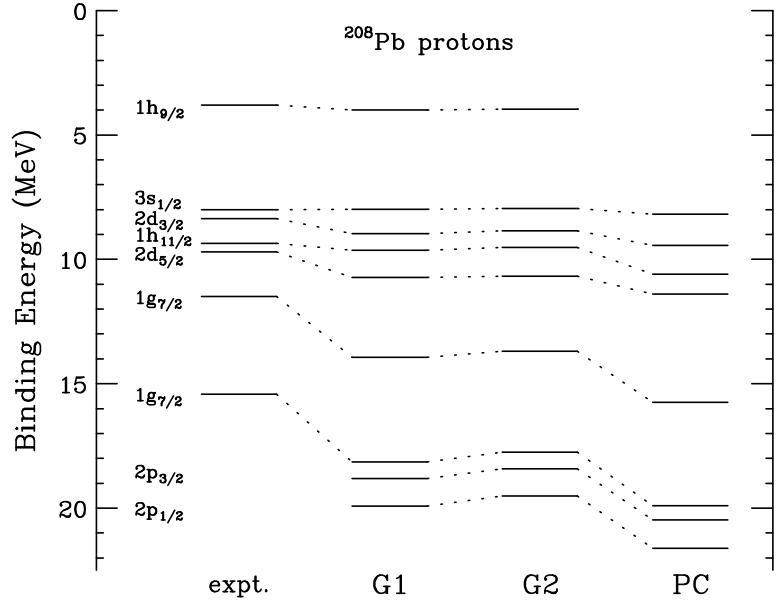}}
\vspace{.2in} \caption{
Predicted proton single-particle spectra for $^{208}$Pb 
near the Fermi surface
using
the G1 and G2 parameter sets from Table~I.
The leftmost values are from experiment \protect\cite{CAMPI72}, 
and PC stands for 
the point-coupling
model of Ref.~\protect\cite{LosAlamos}.
Note that the $1h_{9/2}$ level is the lowest level in 
the first unoccupied shell.}
\label{fig:eleven}
\end{figure}

Our studies of the full $\nu=4$ parametrization yielded  
 two distinct parameter sets (G1 and G2)
with essentially the same $\chi^2$ values.%
\footnote{Small changes in the weights can make the $\chi^2$ lower
for one or the other, but not significantly.}
These sets and several truncated parameter sets described below
 are presented in Table~\ref{tab:one}. 
The observables of the five nuclei
for the G1 and G2 parameter sets are summarized in the tables and figures.
Bulk binding-energy systematics are given in Table~\ref{tab:two}.
The rms charge radii and d.m.s. radii are summarized in 
Table~\ref{tab:three};
we have consistently used the experimental Fourier--Bessel charge 
densities and charge radii of Ref.~\cite{DEVRIES87} for all nuclei.
The spin-orbit splittings of the least-bound nucleons are
given in Table~\ref{tab:four}.
For comparison, we also include results from the point-coupling model
of Ref.~\cite{LosAlamos}. 
Note that our rms radii contains c.m.~corrections,
while those defined in Ref.~\cite{LosAlamos} do not.
The charge densities (and point-proton densities) and charge form
factors for $^{16}$O, $^{40}$Ca,  $^{48}$Ca, and $^{208}$Pb
are shown in Figs.~\ref{fig:one} to \ref{fig:ten}. 
The single-particle spectra for $^{40}$Ca and
$^{208}$Pb are presented in Figs.~\ref{fig:calev} to \ref{fig:eleven}.

\begin{table}[tb]
\caption{Binding-energy systematics for sets G1, G2, and
the point-coupling
model of Ref.~\protect\cite{LosAlamos} (set PC).
Binding energies per nucleon are given in MeV.
}
\vspace{.1in}
\begin{tabular}[tbh]{cccccc}
 Set & $^{16}$O & $^{40}$Ca & $^{48}$Ca & $^{88}$Sr & $^{208}$Pb 
                                          \\ \hline
 G1    &  7.98  &  8.55  &  8.67 &  8.72 &  7.88  \\
 G2    &  7.97  &  8.55  &  8.68 &  8.72 &  7.87  \\
 PC    &  7.97  &  8.58  &  8.69 &  8.75 &  7.86  \\
expt.   &  7.98  &  8.55  &  8.67 &  8.73 &  7.87  \\
\end{tabular}
\label{tab:two}
\end{table}

\begin{table}[tb]
\caption{Rms charge radii (in fm) and d.m.s.\ radii (in fm)
 for sets G1, G2,
and the point-coupling
model of Ref.~\protect\cite{LosAlamos} (set PC).
}
\vspace{.1in}
\begin{tabular}[tbh]{ccccccccccc}
 Set    & \multicolumn{2}{c}{$^{16}$O}   
    & \multicolumn{2}{c}{$^{40}$Ca}  
    & \multicolumn{2}{c}{$^{48}$Ca}  
    & \multicolumn{2}{c}{$^{88}$Sr}  &  
      \multicolumn{2}{c}{$^{208}$Pb} \\ \hline
  &  $\langle r^2\rangle_{\rm ch}^{1\over 2}$ & $R_{\rm dms}$
  &  $\langle r^2\rangle_{\rm ch}^{1\over 2}$ & $R_{\rm dms}$
  &  $\langle r^2\rangle_{\rm ch}^{1\over 2}$ & $R_{\rm dms}$
  &  $\langle r^2\rangle_{\rm ch}^{1\over 2}$ & $R_{\rm dms}$
  &  $\langle r^2\rangle_{\rm ch}^{1\over 2}$ & $R_{\rm dms}$
                               \\ \hline
 G1  & 2.72 & 2.77 & 3.46 & 3.84 & 3.45 & 3.95 
     & 4.19 & 4.99 & 5.50 & 6.80 \\
 G2  & 2.73 & 2.77 & 3.46 & 3.84 & 3.45 & 3.95
     & 4.19 & 5.00 & 5.50 & 6.80 \\
 PC  & 2.73 &       & 3.45 &       & 3.48 &      
     & 4.21 &       & 5.50 &       \\
expt. & 2.74 & 2.76 & 3.45 & 3.85 & 3.45 & 3.96 
     & 4.20 & 4.99 & 5.50 & 6.78 \\
\end{tabular}
\label{tab:three}
\end{table}

\begin{table}[tb]
\caption{Spin-orbit splittings for the least-bound neutrons
and protons  for sets G1,  G2,
and  the point-coupling
model of Ref.~\protect\cite{LosAlamos} (set PC).
}
\vspace{.1in}
\begin{tabular}{c|cccc|cccc} 
\hspace*{.4in}  &
\multicolumn{4}{c}{ neutron $\Delta E_{\rm SO}$ (MeV)}
 &
\multicolumn{4}{c}{ proton $\Delta E_{\rm SO}$ (MeV)}
\\ \hline
  & G1 & G2 & PC & expt. & G1 & G2 & PC & expt.
\\ \hline
$^{16}{\rm O}$   & 6.0 & 6.0     & 6.4  &  6.2
                 & 6.0 & 6.0     & 6.4  &  6.3
\\       
$^{40}{\rm Ca}$  & 6.6 & 6.5     & 6.8  &  6.3
                 & 6.6 & 6.5     & 6.8  &  7.2
\\       
$^{48}{\rm Ca}$  & 5.8 & 5.6     & 5.9  &  3.6
                 & 6.3 & 6.1     & 6.1  &  4.3
\\       \
$^{88}{\rm Sr}$  & 2.1  & 2.0    & 1.9  &  1.5$^a$
                 & 6.2  & 6.0    & 6.1  &  3.5$^a$
\\       
$^{208}{\rm Pb}$ & 0.8 & 0.9   & 0.9 &  0.9
                 & 1.8 & 1.8   & 2.0 &  1.3
\\
\end{tabular}
\label{tab:four}
$^a$Calculated from Bohr and Mottelson\cite{bohr} 
as in Ref.~\cite{LosAlamos}. 
\end{table}

The fits to nuclear charge radii, d.m.s radii, binding energies, 
and spin-orbit splittings are quite good; they are almost all at
the relative accuracy prescribed by the corresponding weights.
A good reproduction of the spin-orbit force in finite nuclei necessarily 
leads to large scalar and vector mean fields in the interiors of nuclei
or in nuclear matter near equilibrium density.
Our calculations show that our parameters produce 
scalar and vector mean fields consistent
with the estimates given in Eqs.~(\ref{eq:fdest}) and
(\ref{eq:derest}). 
The single-particle spectrum in lead is also quite good, although
the level density near the Fermi surface is somewhat too low,
as usual in the Dirac--Hartree
approximation.

The low-momentum behavior of the nuclear 
charge form factors is well reproduced, while the 
high-momentum parts show some departure from the experimental
data.
This is expected, since our approximations are not 
as accurate at higher momentum transfers and we have only constrained
the low-momentum form factor.
In coordinate space,
the vector meson and gradient contributions to the charge densities are
important in reducing the point-nucleon charge fluctuations.
This is evident in the figures, where the point-proton density is
also shown.
We emphasize  that the charge density is obtained directly from the
Hartree solutions; there is no convolution with external single-nucleon
form factors.

\begin{table}[tb]
\caption{Nuclear matter equilibrium properties for sets
from Table~\ref{tab:one} and for the point-coupling
model of Ref.~\protect\cite{LosAlamos} (set PC).
Values are given for the binding energy per nucleon (in MeV),
the Fermi momentum $k_{{\scriptscriptstyle\rm F}}$ (in fm$^{-1}$), the
compression modulus $K$ (in MeV), the bulk symmetry energy coefficient
$a_4$ (in MeV), $M^\ast /M$,
and $\gomega V_0$ (in MeV) at equilibrium.
}
\smallskip
\begin{tabular}[tbh]{cccccccc}
 Set & $E/B-M$ & $k_{{\scriptscriptstyle\rm F}}$
            & $K$ & $a_4$ & $M^\ast/M$ & $\gomega V_0$  \\
 \hline
 W1   & $-16.46$ & 1.279 & 569 & 40.9 & 0.532 & 363 \\
 C1   & $-16.19$ & 1.293 & 304 & 32.0 & 0.657 & 255 \\
 Q1   & $-16.10$ & 1.299 & 242 & 36.4 & 0.597 & 306 \\
 Q2   & $-16.13$ & 1.303 & 279 & 35.2 & 0.614 & 292 \\
 G1   & $-16.14$ & 1.314 & 215 & 38.5 & 0.634 & 274 \\
 G2   & $-16.07$ & 1.315 & 215 & 36.4 & 0.664 & 248 \\
 PC   & $-16.13$ & 1.299 & 264 & 37.0 & 0.575 & 322 \\
\end{tabular}
\label{tab:five}
\end{table}

Experience with a broad class of relativistic mean-field models
shows that models that successfully reproduce bulk properties of
finite
nuclei share characteristic properties in infinite nuclear matter
\cite{FS,bodmer}.
These properties
are the equilibrium binding energy and density, the compression modulus $K$,
the value of $M^\ast/M$ at equilibrium, and the symmetry energy $a_4$.
One further condition is that the light scalar mass satisfies
$500 \lesssim m_{\rm s} \lesssim 550\,{\rm MeV}$.
This condition
ensures reasonably smooth charge densities,
good surface-energy systematics, and an appropriate shape for the 
single-particle potentials \cite{FS}.
This is confirmed from the properties of nuclear matter predicted
from our fits to finite nuclei, as shown in Table~\ref{tab:five} 
for the parameter sets in Table~\ref{tab:one}.

\begin{figure}[t]
 \setlength{\epsfxsize}{4in}
 \centerline{\epsffile{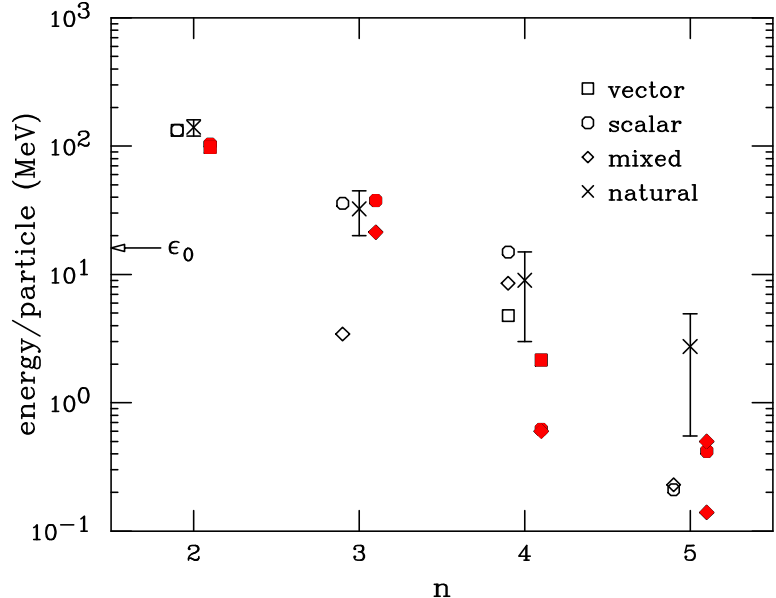}}
\vspace{.2in} \caption{
 Contributions to the energy per particle in nuclear matter for parameter
 sets G1 and G2 from the $n^{\rm th}$-order terms
 of the form $\Phi^l W^m$, where $n = l+m$.  
The boxes are terms with $l=0$, the circles are terms with $m=0$, and
absolute values are shown.
Results from set G1 are open and those from G2 are filled. 
The crosses are estimates based on Eq.~(\ref{eq:generic}).
The arrow indicates the total binding energy $\epsilon_0 =16.1~\mbox{MeV}$.
}
\label{fig:twelve}
\end{figure}

It is instructive to examine the contributions to the energy of
nuclear matter from various terms in the energy density. 
This is shown in Fig.~\ref{fig:twelve}, where the contributions to
the energy per nucleon at equilibrium from
the quadratic, cubic, and quartic powers of the fields $\Phi$ and $W$ in
Eq.~(\ref{eq:nuclmatt}) are
indicated separately.
We set the scale for the natural size of contributions expected from an
$n^{\rm th}$-order term using the power counting estimate 
[Eq.~(\ref{eq:generic})] for
an $n^{\rm th}$-order contact term and the approximation
$\langle\overline NN \rangle \approx
\langle N^\dagger N \rangle = \rhoB$, 
with $\rhoB$ being the baryon density and the bracket indicating an 
expectation value.
These estimates are marked with crosses, and
the error bars reflect a range of $\Lambda$ from 500~MeV to 1~GeV.
We observe that  the estimates are consistent 
with the field energies found from the fits.
Furthermore, these estimates also set the scales for the mean meson fields,
which are consistent with Eq.~(\ref{eq:fdest}).
This is an indication that the naturalness assumption is valid (see below).

Several observations can be made based on the figure: 
\begin{itemize}
 \item
The  contributions decrease steadily with
increasing powers of the fields, so that a truncation is justified.
Note that this decrease will become more gradual as the density
is increased above equilibrium density.
 \item
The nuclear matter binding energy (indicated by the arrow) is an order
of magnitude smaller than the individual $\nu=2$ contributions.
This implies that the $\nu=2$ scalar and vector contributions 
cancel almost completely and also that $\nu=4$ contributions are still
quite significant (at least for set G1).
 \item
The $\nu=5$ contributions [see Eq.~(\ref{eq:fifth})] are representative 
results obtained when the
$\nu=4$ fit is extended. (They correspond to the lowest values of
$\chi^2$).  They are not well determined; almost identical values
of $\chi^2$ can be found for very different $\nu=5$ parameters.  
Nevertheless, even when unnatural $\nu=5$ parameters are introduced, 
their {\it net\/} contribution to the energy is typically a 
few tenths of an MeV.
  Thus a truncation at $\nu=4$ is consistent and adequate.
  \item
The systematics of the two parameter sets at $\nu=4$ are quite
different.  Set G1 parallels the systematics of the standard parameter
sets from Refs.~\cite{RUFA88,REINHARD89,SHARMA93}, 
while the $\nu=4$ terms for G2 are quite small.  This is possible
because of the contributions from the $\alpha_1$ and $\alpha_2$ terms,
which involve gradients of the meson fields 
and thus do not contribute in nuclear matter.
The contributions of these gradients 
[note the signs of the
$\alpha_i$ and the signs in Eq.~(\ref{eq:NLag})]
to the spin-orbit splittings in finite nuclei allow set~G2 
to accurately reproduce the empirical spectra, even though 
$\Mstar/M$ in nuclear matter is larger than in
conventional models.
\end{itemize}

If we return now to the parameters in Table~\ref{tab:one}, we see
 that while the Yukawa couplings and the scalar mass are close
for the two sets, the meson self-couplings are very different.
This is consistent with our earlier Hartree 
calculations \cite{HOROWITZ,bodmer}, which showed that the bulk and
single-particle observables of interest are determined 
to a large extent by a relatively small
number of nuclear properties.
As noted above, 
by reproducing five important properties of nuclear matter and by choosing
an appropriate scalar mass, one essentially fixes the predicted properties
of doubly magic nuclei, and thus the thirteen
parameters in the present model are underdetermined.
The parameters are therefore highly correlated and are sensitive to the
relative weights chosen for the observables, including the spin-orbit
splittings.
To determine the parameters better, one must include additional observables
that are sensitive to different aspects of the dynamics.

To explore these issues further and to put 
our $\nu=4$ fit in perspective, it is useful to find parameter
sets for more constrained truncations.
If we optimize with the ``standard'' mean-field model, 
which includes the Yukawa
couplings, the scalar mass, and $\kappa_3$ and $\kappa_4$, we find
$\chi^2 \approx 110$, and the parameter set 
(which we label Q1) is qualitatively close to
the ``best-fit'' sets of the past and also to the corresponding
parameters of set G1 (see Table~\ref{tab:one}).
Adding the $\nu=4$ vector self-interaction, with parameter $\zeta_0$,
improves the $\chi^2$ to about 90 (set Q2). 
The $\chi^2$ values of sets G1 and G2 are slightly less than 50, so
it is clear that a better fit is obtained with more parameters, 
but not dramatically so.

If we  truncate our model at $\nu=2$ (set W1), 
keeping only the Yukawa
couplings and the scalar mass and setting all other parameters to zero,
the best $\chi^2$ is over 1700.
The ``optimal'' set in this case depends strongly on the relative weights,
because only a subset of the observables can be reproduced accurately.
The standard choice \cite{HOROWITZ}
has been to reproduce the charge radii accurately at the expense of
significant underbinding.  The present set of weights leads to
much better binding energies
(despite a large compressibility)
 but the systematics of the charge radii  
(oxygen in particular) is
badly reproduced.  The spin-orbit splittings are too large in all cases,
corresponding to a low value for $\Mstar/M$ (see Table~\ref{tab:five}). 

At $\nu=3$ (set C1), 
we add two parameters, $\kappa_3$ and $\eta_1$.
(We hold $\eta_\rho$ at zero for simplicity.)
If the number of parameters was
the sole key to a good fit, we would expect $\chi^2$ to be comparable to 
that from the
standard models.  However, the best fit has $\chi^2 \approx 400$.
In practice the $\eta_1$ contribution plays the same role as $\kappa_4$
if one looks at energy contributions in nuclear matter, but it
is insufficient to accurately reproduce nuclear sizes.
It is likely that the density dependence of the two $\nu=3$ terms
is too similar to provide enough flexibility for a good fit.

Thus we find that we need at least $\nu=4$ terms to
find a good fit to nuclear properties.
Examples are sets Q1 and Q2.
Once we allow all $\nu=4$ terms, however, the fit is underdetermined,
as is evident by the differences between our sets G1 and G2.
One of the important new features are the
$\alpha_1$ and $\alpha_2$ parameters, which accompany derivative corrections.
Without these included in the fit, set G2 cannot achieve a $\chi^2$ better
than 100.

If we include  $\nu=5$ parameters from Eq.~(\ref{eq:fifth}) in the fit,
only small improvements in $\chi^2$ ($< 2$ units) are found.
The net contribution to the energy
from these terms is very small (roughly a few tenths of an MeV) 
in these fits, even if the
coefficients are artificially made large so that individual contributions
are significant.
Furthermore, the changes in the lower-order coefficients 
from the $\nu=4$ fits are small.
Thus we find $\nu=4$ contributions are necessary and sufficient for
this set of observables.
Clearly the $\nu=5$ results, when included, do not drive the physics,
and their effects can be absorbed into small changes in the lower-order
couplings.
The insensitivity to the $\nu=5$ 
terms validates  our truncation of the lagrangian.

The parameters in Table~\ref{tab:one} have been displayed in
such a way that they should all be of order unity according to NDA
and the naturalness assumption. This is seen to be the case.%
\footnote{The most unnatural coefficient is $\kappa_4$ from set G1.
Its contribution to the energy is still compatible with natural 
estimates despite its size because of the $1/4!$ counting factor.} 
It is important to emphasize this nontrivial result:  
Without the guidance from NDA, these coefficients could
be any size at all!
We conclude that NDA and the naturalness assumption are indeed compatible 
with and implied by the observed properties of finite nuclei,
even though we are absorbing many-body effects in the coefficients.

However, the detailed mechanism by which the nuclear observables 
constrain the
parameters to be natural is not transparent.
Second-order ($\nu=2$) coefficients vary only slightly, regardless of
the number of parameters used.
The other parameters, while natural, are not as well determined.
We searched for  unnatural parameter
sets by starting from  unnatural values, but 
 the optimization always produced natural results. 
The $\nu=5$ parameters can be made large, but they are
tuned by the optimization so that they cancel almost completely
in the Hartree equations.
Despite the narrow range of densities present in ordinary nuclei,
the different density dependence of various terms in the energy
is apparently sufficient to enforce naturalness.

\section{Discussion and Summary}

Our goals in this paper are to construct an effective field theory
based on hadrons that is
appropriate for calculations of finite-density nuclear systems and to
propose methods for making these calculations tractable.
The lagrangian is consistent with the underlying symmetries of QCD,
and heavy, non-Goldstone bosons are introduced to avoid the calculation
of (some) virtual pion loops.
To organize the lagrangian, we rely on Naive Dimensional Analysis (NDA),
which allows us to extract the dimensional scales of any term, on the
assumption of naturalness, which says that the remaining dimensionless
coefficient for each term should be of order unity, and on the
observation that the ratios of mean meson fields (and their gradients) to
the nucleon mass (which generically represents the ``heavy'' mass scale)
are good expansion parameters.
Since the meson fields are roughly proportional to the nuclear density,
and since the spatial variations in nuclei are determined by the momentum
distributions of the valence nucleon wave functions, this organizational
scheme is essentially an expansion in $k_{\rm F}/M$, where $k_{\rm F}$ is 
a Fermi wavenumber corresponding to normal nuclear densities.
If NDA and the naturalness assumption are valid, one can expand the 
lagrangian in powers of the fields and their derivatives, truncate at some
finite order, and thus have some predictive power for the properties of
nuclei.
Our fits to bulk and single-particle nuclear properties show that this is 
indeed the case.

As in all effective field theories, the coefficients in our nonrenormalizable
lagrangian implicitly include contributions from energy scales heavier
than our generic mass $\Lambda$.
Thus contributions from baryon vacuum loops and non-Goldstone boson loops
(including, in principle, particles more massive than those considered here)
are parametrized in this way.
Moreover, since we presently compute the nuclear observables at the 
one-baryon-loop
or Hartree level, fitting the parameters to nuclei means that they also
implicitly include the effects of many-body corrections.
These consist of nucleon exchange and correlation contributions, which may
involve virtual states with large excitation energies, modifications of
the vacuum dynamics at finite density, like the nuclear Lamb shift, and
virtual pion loops, since the mean pion field vanishes in spherical nuclei.
These many-body effects involve both short- and long-range dynamics,
so it is notable that our fitted parameters are still natural.

We can understand this result with the following arguments.
First, it is unlikely that natural parameters arise from a sensitive
cancellation between unnatural parameters in the lagrangian and compensating
large many-body effects.
It is much more likely that the parameters in the lagrangian are natural,
and the modifications from many-body effects are small.
This is consistent with explicit calculations of relativistic exchange and
correlation effects, which show that these contributions do not significantly
modify the nucleon self-energies nor introduce large state
dependence, at least for states in the Fermi sea.\footnote{%
A well-known, but somewhat exceptional example concerns the symmetry
energy in nuclei.
Explicit calculations show that one-pion exchange graphs make a significant 
contribution to the symmetry energy \protect\cite{DHF}, but these effects
can be simulated at the Hartree level by simply choosing a rho-nucleon
coupling that is larger than what one would expect from free-space
considerations.}
This ``Hartree dominance'' of the mean fields \cite{bodmer}
and of the parameters needed
to produce them is a crucial element in the expansion and truncation scheme
described above.
An important topic for future work is the explicit evaluation of correlation
effects in this model, both to disentangle these contributions from the
fit parameters and to verify the extent to which Hartree dominance is valid.

We can also understand why earlier relativistic Hartree approaches to
nuclear structure were successful.
Many of these calculations used lagrangians that were truncated arbitrarily
at quartic interactions in the meson fields, but we have seen that quintic
and higher-order terms are unimportant (and undetermined)
for the nuclear observables of interest.
Moreover, even though most of these investigations arbitrarily set some of 
the cubic and quartic meson parameters to zero, we have also seen that the
full complement of parameters through fourth order ($\nu \leq 4$) 
are underdetermined by
the data.
Therefore, keeping only a subset of the parameters does not preclude the
possibility of a realistic fit to nuclei.

In summary, we have constructed an effective lagrangian
that maintains the symmetries of QCD, such as  Lorentz invariance,
parity conservation, chiral symmetry, and electromagnetic gauge invariance.
The bulk and single-particle properties of nuclei can be described 
realistically by this effective hadronic theory at the one-baryon-loop level.
Naive dimensional analysis and the naturalness assumption are found to be 
compatible with the observed nuclear properties.
The fitted parameters are natural, and the results are not driven by the 
last terms kept.
The description of the spin-orbit splittings of the least bound
nucleons and the shell structure are generally good, although the errors are
relatively large compared with the other observables.
The long-range electromagnetic structure of the nucleon is included
explicitly using vector-meson dominance and derivative couplings to the
photon; it is not necessary to introduce {\em ad hoc\/} 
electromagnetic form factors.
The mesonic and gradient contributions are important for reducing 
fluctuations in the point-nucleon charge density, and the resulting nuclear
charge form factors agree well with experiment.
The extension of these calculations beyond one-baryon-loop order is
currently in progress.

\acknowledgments

We thank B.~Clark, P.~Ellis, S.~Jeon, L.~McLerran, and J.~Rusnak 
for useful comments and stimulating discussions.
This work was supported in part by the Department of Energy
under Contract No.\ DE--FG02--87ER40365, the 
National Science Foundation
under Grants No.\ PHY--9511923 and PHY--9258270, and the A. P. 
Sloan Foundation.

\end{document}